\documentclass[prd,twocolumn,showpacs,preprintnumbers,amsmath,amssymb,twoside]{revtex4}

\usepackage{graphicx}             
\usepackage{dcolumn}              
\usepackage{bm}

\begin{document}

\preprint{APS/123-QED}

\title{Simulation of Atmospheric Muon and Neutrino Fluxes with CORSIKA}

\author{J\"{u}rgen Wentz}
\email{wentz@ik.fzk.de}
\altaffiliation[\\ Also at ]{Forschungszentrum Karlsruhe, Institut f\"{u}r Kernphysik}
\author{Iliana M. Brancus}
\affiliation{{``}Horia Hulubei'' National Institute of Physics and Nuclear
Engineering, P.O.Box MG 6, 76900 Bucharest, Romania}

\author{Alexandru Bercuci}
\author{Dieter Heck}
\author{J\"{u}rgen Oehlschl\"{a}ger}
\author{Heinigerd Rebel}
\altaffiliation{Also at Fakult\"{a}t f\"{u}r Physik und Astronomie, Universit\"{a}t Heidelberg}
\affiliation{Forschungszentrum Karlsruhe, Institut f\"{u}r Kernphysik,
Postfach 3640, 76021 Karlsruhe, Germany}

\author{Bogdan Vulpescu}
\affiliation{Universit\"{a}t Heidelberg, Physikalisches Institut,
Philosophenweg 12, 69120 Heidelberg, Germany}

\date{\today}

\begin{abstract}
The fluxes of atmospheric muons and neutrinos are calculated by a
three dimensional Monte Carlo simulation with the air shower code CORSIKA
using the hadronic interaction models DPMJET, VENUS, GHEISHA, and UrQMD.
For the simulation of low energy primary particles the original CORSIKA
has been extended by a parametrization of the solar modulation and a microscopic
calculation of the directional dependence of the geomagnetic cut-off functions.
An accurate description for the geography of the Earth has been included by a
digital elevation model, tables for the local magnetic field in the
atmosphere, and various atmospheric models for different
geographic latitudes and annual seasons. CORSIKA is used to calculate 
atmospheric muon fluxes for different locations and the neutrino fluxes for
Kamioka. The results of CORSIKA for the muon fluxes are verified by an
extensive comparison with recent measurements. The obtained neutrino fluxes are
compared with other calculations and the influence of the hadronic interaction
model, the geomagnetic cut-off and the local magnetic field on the neutrino
fluxes is investigated.
\end{abstract}

\pacs{95.85.Ry  
      96.40.Tv} 

\maketitle

\section{Introduction}

Atmospheric neutrinos are produced by the interaction of the primary cosmic
radiation with the Earth's atmosphere. They result 
mainly from the decay of the charged pions and muons:
\begin{eqnarray}
\label{equ:neutrino_sources}
& \pi^+\,\,\, \rightarrow & \mu^+ + \nu_\mu \\
& & \hspace{0.2cm} \hookrightarrow e^+ + \nu_e + \overline{\nu_\mu} \nonumber \\
& \pi^-\,\,\, \rightarrow & \mu^- + \overline{\nu_\mu} \nonumber \\
& & \hspace{0.2cm} \hookrightarrow e^- + \overline{\nu_e} + \nu_\mu, \nonumber
\end{eqnarray}
and to about 10\,\% from similar reaction chains for kaons.
A simple balancing of the different neutrino species involved results in
the following approximate relations between the numbers of neutrinos:
\begin{equation}
\label{equ:neutrino_ratios}
\frac{\,\nu_e\,}{\overline{\nu_e}} = \frac{\mu^+}{\mu^-}, \hspace{0.8cm}
\frac{\,\nu_\mu\,}{\overline{\nu_\mu}} = 1, \hspace{0.3cm}{\rm and} \hspace{0.3cm}
\frac{\nu_\mu + \overline{\nu_\mu}}{\,\nu_e + \overline{\nu_e}\,} = 2.
\end{equation}
A more detailed calculation leads to an energy dependence of all the ratios.
Especially the ratio of muon neutrinos to electron neutrinos strongly
depends on the energy, because the number of muons reaching sea-level before
decaying increases with the energy.

A precise simulation of atmospheric neutrino fluxes is of essential interest
for the interpretation of the so-called atmospheric neutrino anomaly, i.e.
the observation of several neutrino detectors~\cite{sk_1,sk_2,imb,kam,soudan1,
soudan2}  
that the ratio of muon neutrinos to electron neutrinos in the atmosphere differs
approximately by a factor of 2 from the theoretical predictions.
The flux of electron neutrinos seems to agree relatively well
with the expectation, and the anomaly results mainly from a lack of muon
neutrinos. 

In addition, the anomaly displays a pronounced dependence on the angle
of incidence. 
The highest deficit is measured for neutrinos traveling
through the Earth for entering the detectors in upward direction,
while for downward going neutrinos agreement with the theory is found.
This directional dependence of the anomaly is commonly
interpreted in terms of neutrino oscillations. 

Due to the
enormous size of the detector, the results obtained by the Super-Kamiokande
experiment near Kamioka, Japan are statistically most significant and
allow a most detailed exploration of the anomaly. As only detector so far,
Super-Kamiokande could establish a pronounced East-West-effect in the
neutrino flux originating from the influence of the Earth's magnetic field
on the trajectories of the charged primary and secondary cosmic ray
particles~\cite{sk_east_west}. 

\begin{table*}

\vspace{0.2cm}

\caption{\label{tab:models}Features of the different models applied in 
the calculation of atmospheric neutrino fluxes. The following 
abbreviations are used in the table: IGRF stands for the International
Geomagnetic Reference Field~\cite{igrf}, WMM for the World Magnetic Field
Model~\cite{wmm1,wmm2}, and USSA for the US Standard Atmosphere~\cite{ussa}.
The terms used in the table are explained in Sec.~\ref{sec:corsika}.}

\vspace{0.2cm}

\begin{ruledtabular}
\begin{tabular}{ccccccccc}

     & BGS & BN & HKHM & LBK & TKNW & BFLMSR & HKKM & Ply \\

\hline\\

Hadronic interaction & TARGET &    semi-   & FRITIOF/ & TARGET & GEANT & FLUKA  & FRITIOF/NUCRIN & GEANT \\
model                &        & analytical &  NUCRIN  &        &       &        & DPMJET III \\
                     &        &            &          &        &       &        & (parametrized)\\
\vspace{-0.1cm}\\
Dimensions           & 1      &  1         & 1        & 3       & 3     & 3 & 3 & 3\\   
\vspace{-0.1cm}\\
Directional dependence & dipole & dipole- & IGRF & dipole & IGRF & IGRF & dipole & WMM\\
of geomagnetic cut-off &        & like \\ 
\vspace{-0.1cm}\\
Penumbra of cut-off & no & no & no & no & yes & yes & no & yes\\
\vspace{-0.1cm}\\
Local magnetic field & no & no & no & no & yes & no & yes & yes\\
\vspace{-0.1cm}\\
Energy loss by ionization & yes & yes & yes & yes & yes & yes & yes & yes\\
\vspace{-0.1cm}\\
Multiple scattering of muons & no & no & no & no & yes & yes & yes & yes\\
\vspace{-0.1cm}\\
Atmospheric model & USSA & Dorman & USSA & ? & USSA & USSA & USSA & USSA \\
& & model \\
\vspace{-0.1cm}\\
Elevation model of the & no & no & no & no & no & no & no & no \\
Earth\\
\end{tabular}
\end{ruledtabular}
\vspace*{0.2cm}
\end{table*}

The fluxes of atmospheric neutrinos have been calculated with various
theoretical approaches invoking different hadronic interaction models. 
Detailed calculations have been done by
Barr, Gaisser, and Stanev (BGS)~\cite{gaisser88,barr89,agrawal96}; 
Bugaev and Naumov (BN)~\cite{bugaev89};
Honda, Kasahara, Hidaka, and Midorikawa (HKHM)~\cite{honda90,honda95};
Lee, Bludman, and Koh (LBK)~\cite{lee88,lee90},
Tserkovnyak, Komar, Nally, and 
Waltham (TKNW)~\cite{tserkovnyak1999,tserkovnyak2001};
Battistoni, Ferrari, Lipari, Montaruli, Sala, and Rancati 
(BFLMSR)~\cite{battistoni00,battistoni01,battistoni02};
Honda, Kajita, Kasahara, and Midorikawa (HKKM)~\cite{honda01a,honda01b};
and Plyaskin (Ply)~\cite{plyaskin2001}. A recent review of the 
calculations of atmospheric neutrinos can be found in
Ref.~\cite{gaisser_rev}.

The calculation of BGS is a one dimensional Monte Carlo simulation
made in two steps. First, cascades for different primary energies
and zenith angles are simulated, and subsequently, the energy depending
yields of the secondary particles are weighted by the primary spectrum
and the geomagnetic cut-off characteristics for the detector location.
The hadronic interactions are described with TARGET~\cite{target},
a parametrization of accelerator data with special emphasis on energies
around 20\,GeV.  

\renewcommand{\thefootnote}{\alph{footnote}}

The BN calculation is based on a one dimensional semi-classical
integration of the atmospheric cascade equations in straight forward
approximation over the primary spectrum. The hadronic interaction is
described by an analytical parametrization of double differential
inclusive cross-sections based on a compilation of accelerator data.
This approach neglects many details on the nature of the
hadronic interaction~\footnotemark[1].

\footnotetext[1]{Note added in proof: After acceptance for publication we 
got knowledge about new more detailed numerical calculations of Naumov et al. 
(hep-ph/0201310 and references therein) which essentially confirm the 
results of Ref.~\cite{bugaev89}.}

The HKHM calculation is made by using the air shower simulation
code COSMOS~\cite{cosmos} in a one dimensional Monte Carlo simulation. 
For energies above 5\,GeV the hadronic interaction is described in
the frame of FRITIOF version 1.6~\cite{fritiof} with JETSET 6.3~\cite{jetset}.
At lower energies NUCRIN is used~\cite{nucrin}.

The model applied in the LBK calculations is the  model of BGS, but
extending the calculation on three dimensions. The same primary spectrum
has been used, too. The calculation intends to study the influence of the
transversal momenta in the different reactions on the neutrino fluxes.  

The three dimensional calculation of TKNW is based on
the GEANT 3.21 detector simulation tool~\cite{geant} and its various models
for the  hadronic interaction, CALOR~\cite{calor63,calor70,calor72},
FLUKA92~\cite{fluka92a,fluka92b}, and GHEISHA~\cite{gheisha}.

Both, the LBK and the TKNW calculation failed to discover
a major enhancement of the neutrino fluxes near the horizon, which was 
predicted for the first time
in the three dimensional simulation of BFLMSR. In the meantime, the TKNW 
group revised its model, and finds also an enhancement at the 
horizon~\cite{tserkovnyak2001}.

In the calculations of BFLMSR the FLUKA98 and
FLUKA2000~\cite{fluka01a,fluka01b} are used as models for the hadronic
interaction. These
versions of FLUKA are quite different from the FLUKA92 version integrated
in the GEANT package and being used in the TKNW calculation. 

HKKM extended the calculation of HKHM to three dimensions. Additionally, the
interaction models of COSMOS can be 
replaced now by a parametrized version of DPMJET III~\cite{dpmjetIII}. Meaning
that instead of interfacing DPMJET to COSMOS, DPMJET is run at fixed
energies and the yields of secondary particles are parametrized. This enhances
very much the calculation speed, but subtile details of the interaction
model might be lost in this approach. The published results being used in
this paper for comparisons are based on the hadronic interaction models of the
original COSMOS.

The calculation of Plyaskin is based also on the GEANT detector simulation
package, but only the GHEISHA model is used for the simulation of the hadronic
interaction. The atmosphere is sampled in layers with constant density
of 1\,km thickness.

The major differences of the various neutrino calculations 
in handling certain physical effects and the geographical details of
the Earth are compiled in Tab.~\ref{tab:models}. 

In this communication a full three dimensional simulation procedure for
atmospheric muon and neutrino fluxes is presented using the standard air shower
simulation code CORSIKA~\cite{corsika}. In contrast to the previous
calculations which assumed the Earth for instance as mathematical sphere, 
the actual attempt
includes a complete description of the geographical parameters of the Earth.
For this purpose the CORSIKA 6.0 code is extended by a precise calculation of the
geomagnetic cut-off, a parametrization of the solar modulation, a digital
elevation model of the Earth, tables for the local magnetic field in the
atmosphere, and various atmospheric models for different climatic zones and
annual seasons.

As evident from Equs.~\ref{equ:neutrino_sources} and \ref{equ:neutrino_ratios}
the correlation between neutrinos and muons is very direct, especially the
charge ratio of muons reflects the ratio of electron neutrinos to
electron antineutrinos. Thus, the calculation results of CORSIKA are
verified by the simulation of atmospheric muons and their comparison with
recent measurements.

The procedure of simulation is demonstrated by a detailed calculation
of atmospheric neutrino fluxes in Kamioka.
Using the versatility of the CORSIKA program to cooperate with different models
for the simulation of the hadronic interaction, special emphasis is
put on the question, how various formulations of the hadronic interaction
influence the fluxes of atmospheric neutrinos. It will be shown that
uncertainties in the description of the hadronic interaction are the
main error source for the calculation of atmospheric neutrino fluxes.

Furthermore, the effects of the geomagnetic cut-off modulating the primary
flux and of the local magnetic field which deflects the charged shower
particles on their way through the atmosphere are studied in detail.
Repeating the calculations, setting once the geomagnetic cut-off 
and once the local magnetic field zero, allows to disentangle the 
individual influences. 

It will be proven, that the local magnetic field
is by far not negligible and leads to an increase of the ratio of electron
neutrinos to electron antineutrinos. Furthermore, it modulates the
azimuthal dependence of the neutrino fluxes and causes an East-West-effect,
clearly visible for sites with a low geomagnetic cut-off.

\section{The simulation tool CORSIKA and its extensions for the
simulation of low energy atmospheric particles}
\label{sec:corsika}

\subsection{The air shower simulation program CORSIKA}

The simulation tool CORSIKA has been originally designed for the
four dimensional simulation of extensive air showers with primary
energies around $10^{15}$\,eV. The particle
transport includes the particle ranges defined by the life time of
the particle and its cross-section with air. The density profile of
the atmosphere is handled as continuous function, thus not sampled
in layers of constant density. 

Ionization losses, multiple
scattering, and the deflection in the local magnetic field
are considered. The decay of particles is simulated in exact
kinematics, and the muon polarization is taken into account.

In contrast to other air shower simulations tools, CORSIKA offers
alternatively six different models for the description of the high energy
hadronic interaction and three different models for the description
of the low energy hadronic interaction. The threshold
between the high and low energy models is set by default to
$E_{Lab} = 80$\,GeV/n.
 
Due to the steep spectrum of primary cosmic rays, only some 10\,\% 
of the neutrinos detected in the Super-Kamiokande experiment originate from 
primary particles with energies higher than 80\,GeV/n and the
quality of the simulated neutrino fluxes mainly depends on the
trustiness in the models describing the low energy hadronic
interaction.   

Nevertheless, the extent to which
different high energy interaction models are able to reproduce
experimental muon data has been investigated in a previous
paper~\cite{hh_wentz}. It has been shown that
DPMJET II.5~\cite{dpmjet1,dpmjet2,dpmjet3} and VENUS 4.125~\cite{venus}
agree best with muon data, while QGSJET~\cite{qgsjet} and 
SIBYLL 1.6~\cite{sibyll1,sibyll2} are not reproducing well the charge 
ratio of muons above 80\,GeV/n. Thus, in this paper only DPMJET and 
VENUS are used for the simulation of the high energy hadronic
interaction.

For the simulation of the low
energy hadronic interaction GHEISHA~\cite{gheisha} and UrQMD 1.1
~\cite{urqmd1,urqmd2} are applied.
Additionally DPMJET includes some extensions which allow also the simulation
of the hadronic interaction down to energies of 1\,GeV. In this case 
UrQMD is used for the simulation of hadronic interactions with 
energies below. The total number of muons and neutrinos resulting from 
hadronic interactions below 1\,GeV is very small, thus UrQMD plays more
the role of a technical fallback, to prevent the program from
crashing. No real influence of UrQMD is noticeable in the physical
results in this case.

Low energy reactions are handled very similarly in DPMJET~II.5
and DPMJET~III, so that the results obtained with both versions in the
energy range relevant for the atmospheric neutrino anomaly should be
fully comparable.

Fluxes calculated by CORSIKA have statistical errors, caused
by the limited number of particles calculated in the Monte Carlo simulation,
and various systematic errors. It can be assumed that the main sources of systematic
errors  result from the primary spectrum and the hadronic interaction models. 
Errors due to the particle tracking or particle decay are hardly to be
quantified but they should be negligible compared to the other error
sources. All errors given in the further results are purely statistical.

\subsection{The fluxes of primary cosmic particles} 
\label{sec:prim_flux}

\begin{figure}
\includegraphics[width=8.5cm]{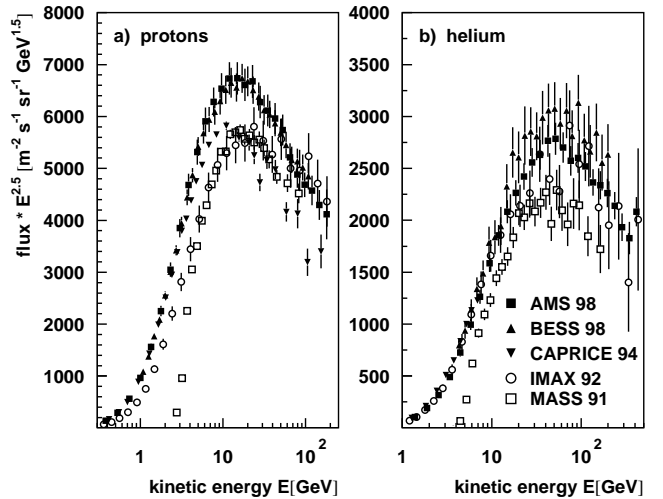}
\caption{\label{fig:prim_flux} The fluxes of primary protons (a) and
helium nuclei (b) as measured by recent balloon and satellite borne 
experiments. In order to enhance the differences in the region of
interest, the fluxes are multiplied by $E^{2.5}$.}
\end{figure}

A major uncertainty in the early calculations of atmospheric
particle fluxes stems from the absolute primary particle fluxes.
These are measured by satellite or balloon borne experiments,
operating above or at the limit of the Earth's atmosphere. In 
Fig.~\ref{fig:prim_flux} the results of recent experiments are 
compiled.
The balloon experiment MASS~\cite{mass_p} has been operated in Fort Sumner,
New Mexico, where the vertical geomagnetic cut-off rigidity is 4.2\,GV,
explaining the missing flux below the cut-off. The balloon experiments
BESS~\cite{bess_p}, CAPRICE~\cite{caprice_p},
and IMAX~\cite{imax_p} were launched in Lynn Lake, Canada, near the
geomagnetic pole with a very low cut-off rigidity of about 0.5\,GV.
The Space Shuttle mission of the
AMS prototype~\cite{ams_p,ams_he} collected data over a large range of cut-off
rigidities ranging from the maximum rigidity at the geomagnetic 
equator down to vertical cut-off rigidities less than 0.2\,GV, corresponding
to proton momenta well below the pion production threshold.

At low energies, especially below 10\,GeV, solar modulation becomes 
important and introduces a further, time dependent source of differences
between the data. But the experiments differ also at higher energies. 
The results of AMS and BESS agree perfectly within experimental errors
while all other experiments report fluxes, being mostly about 15 - 20\,\% 
lower. The differences between the experiments are not constant in energy 
and can therefore not be explained by a simple offset in the 
energy calibration. For instance the MASS and IMAX results agree at
higher energies with the AMS data while the CAPRICE results match at 
lower energies.

AMS and BESS detectors have been calibrated at accelerator beams
of protons (BESS, AMS), He and C nuclei (AMS). This ensures that the
performance of the detectors and the analyzing procedure were thoroughly
understood, giving high evidence that the higher primary proton fluxes 
reported by AMS and BESS are the better ones.

The results for primary helium nuclei show similar differences between
the experiments like the results for primary protons. Again AMS and  
BESS report higher fluxes than the other experiments, but the results 
of AMS and BESS do not agree completely. The cross calibration with light
ions in case of AMS is a strong argument for the correctness of the AMS 
data.

Different from the primary flux parametrization, proposed recently
by Gaisser et al.~\cite{gaisser_prim_flux},
where the helium flux is obtained by a combined fit of the AMS and BESS
results, the calculations in this paper are based on the AMS results,
only. The primary particle generator in CORSIKA uses power laws
extracted from the higher energy data of AMS including
the solar modulation and the geomagnetic cut-off as described in 
the next sections.

The bulk of primary particles producing neutrinos with energies
being detected in Super-Kamiokande is covered by the momentum
acceptance of AMS. In order to avoid any artificial cut for higher
energies, the power laws have been just extrapolated up
to the knee region. As our knowledge of the cosmic radiation at 
higher energies is rather poor, this assumption is still in fair
agreement with the measurements.

\subsection{The description of the solar modulation}

The sun emits a magnetized plasma with a velocity of 
100 - 200\,km/s~\cite{solar_mod}. To reach the Earth,
galactic cosmic rays have to diffuse into the inner  
heliosphere against the outward flow of the turbulent
solar wind, a process know as solar modulation. Depending on
the solar activity the lowest energy cosmic particles 
reach the Earth with a variable flux.

For most places on Earth the geomagnetic cut-off alters the
primary particle fluxes
stronger than the influence of the solar
modulation. Therefore the geomagnetic cut-off must be
simulated in a detailed microscopic calculation as described 
in Sec.~\ref{sec:cut_off} while the solar modulation
can be handled by the parametrization of Gleeson and
Axford~\cite{gleeson}. This parametrization is based on a spherical
symmetrical model in which the differential intensity $J(r, E, t)$ 
for the total energy $E$ in the distance $r$ from the sun for the
time $t$ is given by
\begin{equation}
J(r, E, t) = \frac{E^2 - E^2_0}{\lbrack E+\psi(t) \rbrack ^2 - E^2_0}
\hspace{0.2cm} J \lbrack \infty,E+\psi(t) \rbrack ,
\label{equ:solar_mod}
\end{equation}
with $E_0$ being the rest mass and $\psi(t)$ is a free, time depending
parameter which can be interpreted as the energy loss of a primary
particle during its approach to the Earth. 

\begin{figure}
\includegraphics[width=8.3cm]{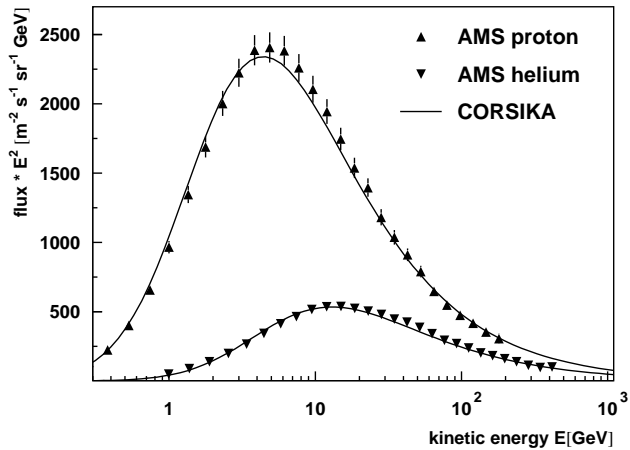}
\caption{\label{fig:prim_fit} The fluxes of primary protons and
helium nuclei as obtained by the primary particle generator
of CORSIKA including the solar modulation but no geomagnetic 
cut-off compared with the results of AMS. In order to pronounce
the differences in the region of
interest, the fluxes are multiplied by $E^{2}$.}
\end{figure}

In principle, $\psi(t)$ can be deduced within theoretical models from
the solar activity.
Nevertheless for the calculation of the neutrino fluxes in Kamioka,
$\psi(t)$ can be assumed as constant in time. 
The flight of AMS took place roughly in the middle of the data taking
period of Super-Kamiokande, thus the values of $\psi$ for primary
protons and helium nuclei are obtained directly by a fit of the function
in Equ.~\ref{equ:solar_mod} to the low energy part of the spectra measured 
by the AMS experiment. The resulting absolute primary particle spectra
without considering the geomagnetic cut-off used for the primary particle
generator of CORSIKA are shown in Fig.~\ref{fig:prim_fit}. 
The overall agreement is quite good, but a systematical 
deviation around 10\,GeV indicates that the used parametrization
is not the best possible. Nevertheless the deviation remains mostly
within the experimental errors and the highest discrepancy for a 
single point is found as 6\,\%. The additional error caused by this
in the atmospheric particle fluxes is quite small.

\begin{figure}
\includegraphics[width=8.5cm]{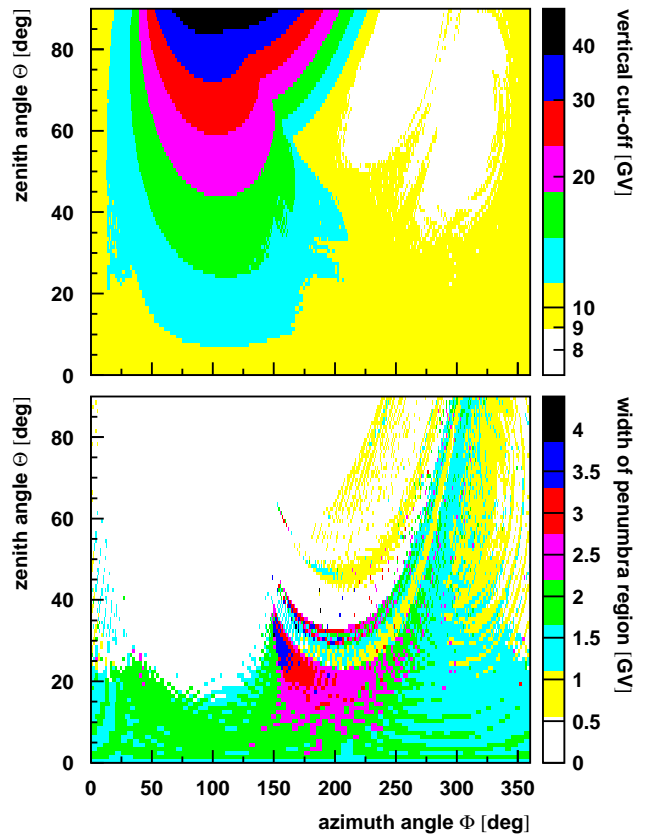}
\caption{\label{fig:kam_cutoff} The mean geomagnetic cut-off and the width
of the penumbra region for Kamioka, Japan. The width of the penumbra
region is defined by the rigidity difference of the lowest momentum of an
antiproton escaping to outer space and the highest momentum of an antiproton being
trapped.
The counting of the azimuth angle $\phi$, here and in all further plots,
follows the convention used by the Super-Kamiokande detector,
$\phi$ = 0$^\circ$ means looking to the South (the particle travels to the North),
$\phi$ = 90$^\circ$ to the East (the particle travels to the West), etc.
The North direction here is defined as the geographical one. The angle
between the geomagnetic and geographic North direction in Kamioka
is -7.59$^\circ$.
}
\end{figure}

\subsection{The simulation of the geomagnetic cut-off}
\label{sec:cut_off}

The Earth's magnetic field has nearly the shape of
a dipole field. The field is strong enough to deflect charged primary particles
on their way to the Earth's surface. While near the geomagnetic poles 
particles with very low momenta can penetrate to the Earth's surface,
protons with up to 60\,GeV, impinging horizontally near the geomagnetic 
equator are reflected back to space.

The calculation of the geomagnetic cut-off is done in a Monte Carlo
simulation of the possible particle trajectories in the so-called
back-tracking method. Instead of tracking primary protons
from outer space to the Earth's surface, antiprotons from the
surface are retraced to outer space.
This method has the advantage that it allows to calculate in a 
straight forward way a table of allowed and forbidden trajectories.
The entries in the table depend on the location on Earth, the
arrival direction and the particle momentum.

In detail the particle tracking starts at 112.83\,km, the top of atmosphere
as defined in CORSIKA. The influence of the local magnetic field in
the atmosphere, including the deflection of charged shower particles
is handled lateron by CORSIKA using the approximation of a homogenous
field.

The particle tracking is based on GEANT 3.21~\cite{geant} and the magnetic
field is described by the International Geomagnetic Reference
Field~\cite{igrf} for the year 2000. For the downward going particle
fluxes the location where the primary particle enters the atmosphere
is bound to the nearer surrounding of the experiment and the arrival
direction is sampled in cells of a solid angle of 250 $\mu$sr. For upward
going neutrinos the geomagnetic cut-off is calculated for 
1655 locations, distributed nearly equidistant over the Earth's
surface, and the angle of incidence for each location is sampled in cells
of 48\,msr. 

Instead of calculating a sharp cut-off, functions in momentum steps of
0.2\,GeV/c up to a maximum momentum of 64\,GeV/c are evaluated. This procedure
accounts for the penumbra region of the cut-off, i.e. the chaotic change
from open and closed trajectories which can be observed in irregular magnetic
fields, as in case of the geomagnetic field.

As an example for the results
obtained for a fixed detector location, the mean geomagnetic cut-off
for particles entering the atmosphere in Kamioka is shown
in Fig.~\ref{fig:kam_cutoff}.
Local irregularities of the magnetic field over Japan cause a
remarkable strong deviation from the regular shape expected 
for a magnetic dipole field.
Assuming highly accurate Monte Carlo simulations
and highly accurate measurements, this feature should be reflected in
the zenithal and azimuthal dependence of the particle intensities
in Kamioka.

Kamioka has a very extended penumbra region which exceeds a width of 
4\,GV in some particular directions. Details about the simulation of
the geomagnetic cut-off and plots for other locations on the Earth
may be found in Ref.~\cite{cutoff_hh}. 

A check of the primary particle generator in CORSIKA with its 
assumptions for the solar modulation and the geomagnetic cut-off can 
be done by the recent results of the AMS-prototype mission~\cite{ams_near_p}. 
Due to the inclination of 51.7$^\circ$ of the shuttle orbit, the space
craft passes geomagnetic latitudes from 0 to more than 1\,rad.

The experimental spectra of downward going protons and
helium nuclei can be compared rather directly with the
results of the primary particle generator. 
Only a correction for the altitude dependence of the geomagnetic
cut-off has to be applied. The cut-off has generally the highest value
at the surface of the Earth, decreases with the altitude and vanishes
when leaving the Earth's magnetosphere. The mean difference in the cut-off
between the top of atmosphere as assumed in CORSIKA and the orbit of the
space shuttle is evaluated by a dedicated GEANT simulation and
has a value of about 10\,\%.

\begin{figure}
\includegraphics[width=7.9cm]{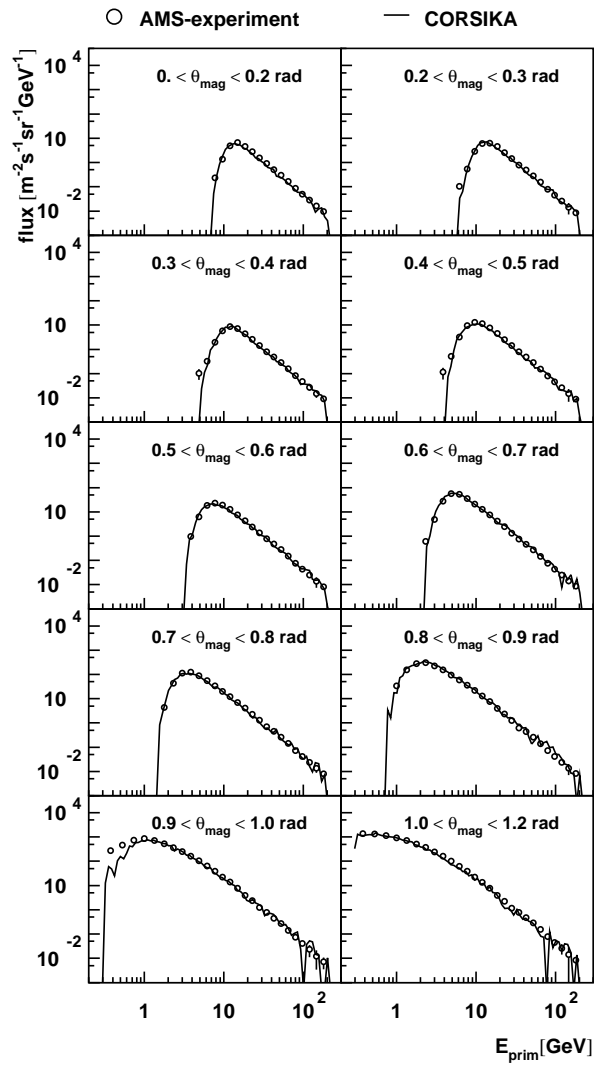}
\caption{\label{fig:ams_cutoff} Comparison of the AMS-results on
downward going primary protons for different intervals of the
geomagnetic latitude with spectra produced by the primary particle
generator of CORSIKA including the simulation of the
geomagnetic cut-off and the parametrization for the solar
modulation.}
\end{figure}

The spectra of primary protons for different
regions of the geomagnetic latitude together with the 
spectra produced by the primary particle generator of CORSIKA are 
shown in Fig.~\ref{fig:ams_cutoff}. 
The agreement between experiment and simulation is very good
and the systematic decrease of the geomagnetic cut-off with the 
geomagnetic latitude is reproduced nicely. Only the spectrum for 
geomagnetic latitudes $0.9 < \theta_{mag} < 1$ shows a noticeable 
difference, which has to be attributed to the low absolute value 
of the cut-off which becomes comparable to the momentum steps used
in the simulation of the cut-off functions. This disagreement has no 
significance for the calculation of atmospheric muon or neutrino
fluxes, because the primary energies are already near or below the
pion  production threshold. The results obtained for primary helium
nuclei have a similar quality.

Particles stored for longer times in the geomagnetic field, the
so-called albedo or sub-threshold particles are not considered
in the present calculations. It has been demonstrated in
Ref.~\cite{lipari_albedo} that they contribute to the atmospheric
particle flux only negligibly. 

\subsection{The geography of the Earth in CORSIKA}
\label{sec:geography}

The geography of the Earth plays a certain role in the simulation
of atmospheric particle fluxes, because the apparent thickness of the 
atmosphere is altered by the different elevation of the terrain over
sea-level and various climatic conditions on the density structure of the
atmosphere. Also the local geomagnetic field, bending charged secondary
particles in the atmosphere has quite a different strength for locations
near the geomagnetic poles and the equatorial regions. For the geomagnetic
poles the absolute field is found to be 64.6\,$\mu$T, while the strength
at the geomagnetic equator is only 21.7\,$\mu$T.

Due to the vicinity of the place where the primary particle enters
the atmosphere and the place of detection in the simulation of vertical
downward going neutrinos, the geographic data are assumed as constant in
the corresponding calculations. For the simulation of inclined particles, 
the distance in the locations may reach already 1200\,km, and in case of 
upward going neutrinos the origin of the primary cosmic particles is
distributed over the entire  Earth.

\begin{figure}
\includegraphics[width=8.5cm]{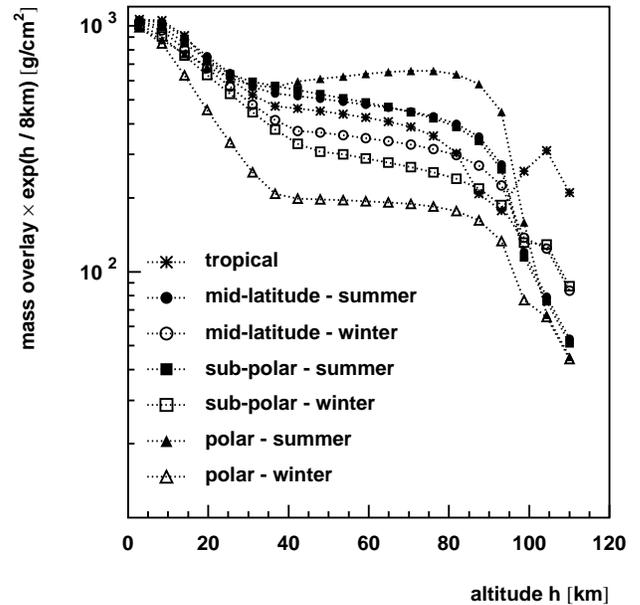}
\caption{\label{fig:atmospheres} The density distribution for atmospheric
conditions of different climatic zones and seasons,
plotted as mass overlay for a given altitude. In order to
enhance the differences, the mass overlay for the altitude $h$ is
multiplied by $\exp(h/8\,{\rm km})$. The differences for altitudes between
20 and 80\,km are most important for atmospheric particle fluxes,
while the differences over 80 km are artificially introduced
by constraining all models to have zero density at 112.83\,km (the starting
altitude of CORSIKA). For the air shower  
development this is negligible, because the mass overlay at 80\,km
amounts to less than $10^{-2}$\,g/cm$^2$.
}
\end{figure}

Therefore, the local geomagnetic field is tabulated on basis of the
International Geomagnetic Reference Field~\cite{igrf} by a table containing
the field parameters for 64800 locations distributed over the Earth's surface. 
The elevation over sea-level is described in a table of equivalent
resolution of data published by the US National Geophysical
Data Center in Ref.~\cite{elevation_model}. 

The atmospheric profiles observed in tropical and polar regions
show considerable differences. The non-tropic atmospheres are subject to
additional variations with the annual seasons. 
The extended CORSIKA code accounts for these effects by 7 atmospheric models
~\cite{atmospheres}. The corresponding density distributions are plotted in 
Fig.~\ref{fig:atmospheres}.
As expected, the largest differences appear between the polar
winter and summer. The seasonal variations become less important and
vanish as the climatic zone approaches the equator.
   
\subsection{The settings and the way of simulation in CORSIKA}

The simulations discussed in this paper have been made using 
the CORSIKA program in version 6.000. All bugs found in CORSIKA until
version 6.014 have been corrected also in the extended version.

The simulation of atmospheric particle fluxes with CORSIKA starts by
selecting the type of primary particles and the ranges for the primary energy, 
zenith and azimuth angles, and by fixing the geographical location on Earth.
The primary energies vary for all simulations reported in this
paper between the minimum geomagnetic cut-off and $10^{15}$\,eV. 

The standard CORSIKA version makes use of a planar atmospheric model. This
is a good approximation as long the zenith angle $\theta$ of the particles 
does not exceed 70$^\circ$. The planar atmosphere approximation is used in
this paper for the calculation of vertical muon fluxes, because the experiments
are usually limited to muons having zenith angles less than 30$^\circ$.
   
For the simulation of the East-West-effect of atmospheric muons and for
all simulations of atmospheric neutrinos the zenith angles must be
varied over the complete range. These simulations have 
been made with the so-called ``curved'' version of CORSIKA.
Here the curvature of the Earth's atmosphere is approximated by sliding
and tilting plane atmospheres. Each time the horizontal displacement of a particle
exceeds a limit of 6 to 20 km (dependent on the altitude), a transition
to a new local plane atmosphere is performed~\cite{schroeder}.

The different primary particles, i.e. protons and helium nuclei are simulated
in separate runs and the ratio between them follows the absolute fluxes
reported by the AMS prototype mission. In order to account for heavier primary 
particles the equivalent number of primary helium nuclei is used. The
absolute fluxes of heavier nuclei are taken from the compilation of
Wiebel-Sooth et al.~\cite{wiebel-sooth}. A justification of this
simplification is provided by the fact that all heavier particles 
contribute together less than 5\,\% to the neutrino flux and all nuclei
have a similar ratio of protons to neutrons.
  
The air shower calculation starts by getting a random location on
the Earth, a random energy and a random arrival direction. If the particle
does not exceed the geomagnetic cut-off for the given location or the solar
modulation, a new set of geographic coordinates, energy and arrival angles
is used. If the particle fulfills the requirements, the geomagnetic parameters,
the altitude, and the atmosphere, are set according to the
geographical position. Due to the long measuring time of Super-Kamiokande,
atmospheric models for summer and winter are used in equal parts. 

The primary particle is tracked
to the first interaction point, given by the cross-section of the particle 
with air. The nuclear reaction is handled by the selected hadronic interaction
model and all secondary particles are tracked up to their decay or further 
interactions.

The obtained numbers of atmospheric particles have to be normalized to the
fluxes of primary particles. For sake of simplicity the number of 
primary particles with an energy larger than 1000\,GeV in
the simulation, being free of any influence of the geomagnetic cut-off and
the solar modulation, is set equal to the integral flux above 1000\,GeV as
extrapolated in Sec.~\ref{sec:prim_flux}. In cases with a limited statistical
accuracy the calibration is made at 100\,GeV. The fluxes at this energy are already
influenced by the solar modulation by some 4.5\,\%, what has
to be taken into account.
   
Due to the flat or partially flat geometry applied in CORSIKA, the obtained 
neutrino fluxes have to be scaled by the surface
difference of the two shells having the radius of the Earth and the
radius of the Earth plus 112.83\,km. This correction leads to a factor of 
1.036.

\section{Calculation of atmospheric muon fluxes}

\subsection{The differential muon flux}

The calculation of atmospheric muon fluxes controls the calculations of
atmospheric neutrino fluxes. The charge ratio of muons provides additional
and partly complementary information. 

Atmospheric muons have been measured over several decades. The data are
compiled in two recent papers~\cite{vulpescu98,hebbeker02} and in the
new review~\cite{grieder}, showing relatively large discrepancies between
the experiments. 
The comparisons of this communication are focused on the recent
measurements of BESS, CAPRICE, the OKAYAMA cosmic ray telescope and
WILLI. In case of BESS~\cite{bess_muon01,bess_muon02} and 
CAPRICE~\cite{kremer99} the results of atmospheric muons
have been obtained in ground based runs, performed as test of the
detectors. 
The OKAYAMA telescope~\cite{tsuji98} is a classical magnetic
spectrometer and WILLI~\cite{vulpescu98,vulpescu01} represents a compact
scintillator experiment dedicated to the precise measurement of the
muon charge ratio. The charge ratio is deduced hereby from the 
different life time of positive and negative muons in matter.

\begin{table}
\caption{\label{tab:geo_params} The geographical parameters for the
different detector sites. The quantity $h$ is the elevation over sea-level,
$R_c$ the mean vertical geomagnetic cut-off, $B_x$ the horizontal
component of the magnetic field, $B_z$ the vertical downward component and
$B_\alpha$ is the angle between the magnetic and geographic north 
direction. The parameters of the magnetic field are valid for the
year 2000 and an altitude of 56.4\,km.  
}

\begin{ruledtabular}
\begin{tabular}{cccccc}
site & $h$ [m] & $R_c$ [GV] & $B_x$ [$\mu$T] & $B_z$ [$\mu$T] & $B_\alpha$ [deg]\\
\hline\\
Bucharest   & 85.   & 5.6  & 21.98 & 40.96 & 3.64 \\
Fort Sumner & 1270. & 4.2  & 22.89 & 44.33 & 9.44 \\
Lynn Lake   & 360.  & 0.5  & 9.81  & 57.51 & 9.36 \\
Okayama     & 5.3   & 11.8 & 30.48 & 34.31 & -6.64 \\
Tsukuba     & 30.   & 11.5 & 29.08 & 34.82 & -6.95 \\
\end{tabular}
\end{ruledtabular}
\end{table}

For the simulation of the atmospheric muon fluxes the precise geographical
parameters, like the geomagnetic cut-off and the altitude of the different
detector sites are taken into account. The used parameters are compiled
in Tab.~\ref{tab:geo_params}. Due to the geographic vicinity
of Okayama and Tsukuba and the same altitude of both sites, the results of the
OKAYAMA telescope can be compared directly with the measurements and
calculations done for Tsukuba.

\begin{figure}[!hb]
\includegraphics[width=8.5cm]{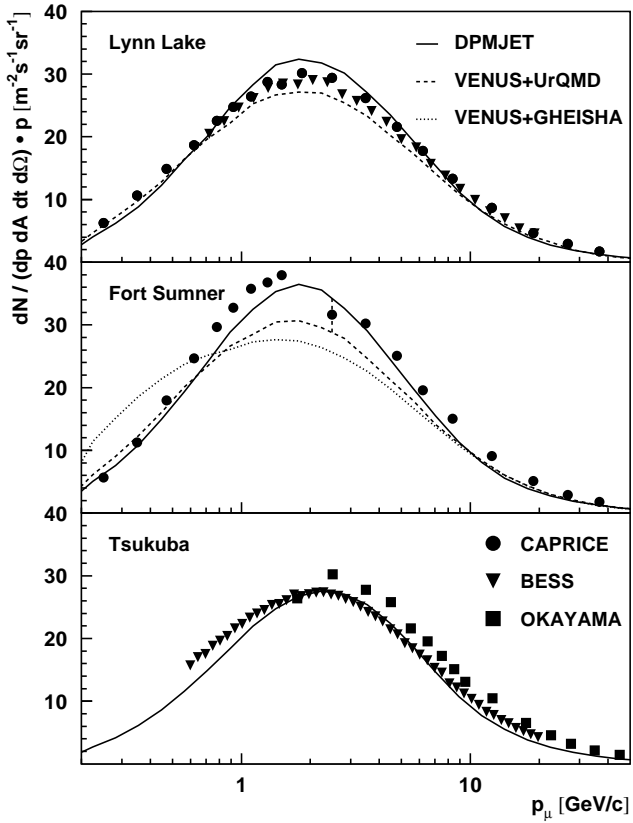}
\caption{\label{fig:cor_diff} The differential flux of vertical muons
calculated by CORSIKA using different models for the description
of the hadronic interaction in comparison with experimental results
for various detector sites. In order to enhance the differences in the 
region of interest, the fluxes are multiplied with the muon momentum $p_\mu$.}
\end{figure}

The results for the differential flux of vertical muons are compiled
in Fig.~\ref{fig:cor_diff}. The calculation with DPMJET as well as the
calculations with VENUS + UrQMD agree generally well with the
experimental data. Only the GHEISHA results show a strange enhancement of the
differential muon flux for low energies and quite a different momentum
dependence.

\subsection{The charge ratio of muons}

\begin{figure}[!ht]
\includegraphics[width=8.5cm]{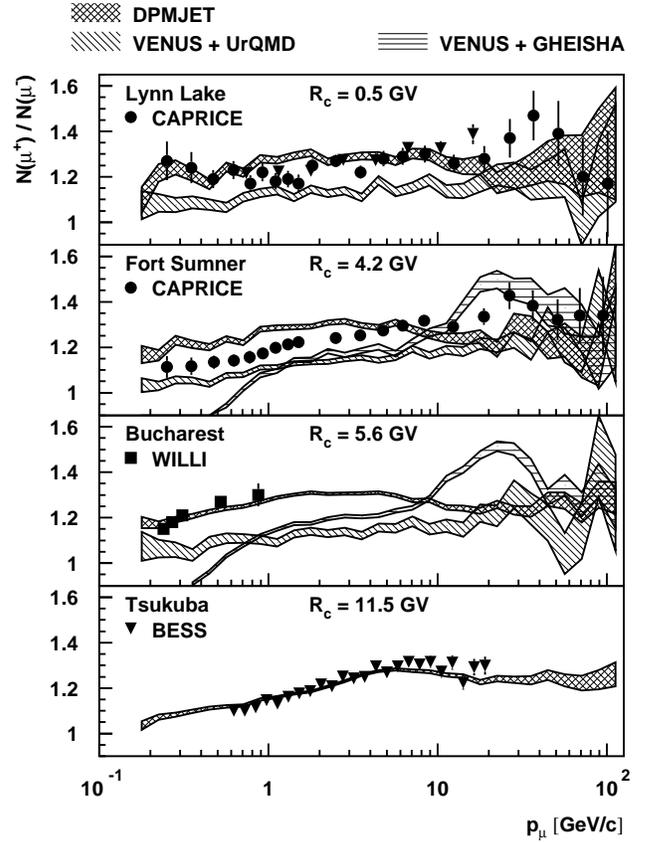}
\caption{\label{fig:cor_charge} The charge ratio of muons
calculated by CORSIKA using different models for the description
of the hadronic interaction compared to experimental data
from various detectors.}
\end{figure}

In contrast to the differential muon fluxes, the charge ratio of muons
reveals larger discrepancies. The CORSIKA results for the charge ratio of
muons are compared in Fig.~\ref{fig:cor_charge} with the experimental data. 
Again the results obtained with the GHEISHA model are far from the 
experimental observations but
there are also differences between the results of DPMJET and VENUS + UrQMD.
The results obtained with VENUS + UrQMD are lower than the experimental
values especially for low and intermediate energies. It has been shown,
that this deviation originates mainly from UrQMD while at higher energies
VENUS leads to a muon charge ratio which is compatible to the
measurements~\cite{wentz01}. 

The DPMJET results agree generally well with the data, with exception 
of the CAPRICE results in Fort Sumner.
The deviation for Fort Sumner has to be questioned because the
geomagnetic cut-off in Fort Sumner resembles that in Bucharest.
Therefore the differences in the experimental values and the continuous
increase of the charge ratio in the CAPRICE measurement for Fort Sumner
far beyond the geomagnetic cut-off seem to indicate experimental
problems in this particular measurement.

The real influence of the geomagnetic cut-off on the muon charge ratio
can be seen when comparing the CAPRICE and BESS results for Lynn
Lake, the WILLI results for Bucharest, and the BESS
results for Tsukuba.
At higher energies the ratio stays nearly constant, however
it decreases, when the geomagnetic cut-off clips the high excess of 
low energy primary protons, as can be observed in the results for 
Bucharest and Tsukuba. This effect
is nicely reproduced by CORSIKA using DPMJET as interaction model, while
using UrQMD the effect is covered by intrinsic problems of the model.

The systematics of the geomagnetic cut-off shows again the problem
of the CAPRICE results for Fort Sumner. The CAPRICE results have practically
the same dependence on the momentum as the BESS results in Tsukuba, where
the geomagnetic cut-off is nearly 3 times higher.

It could be argued that Fort Sumner has an altitude of 1230\,m above sea level
and there could be a strong dependence of the charge ratio on the altitude,
but the CORSIKA simulations include the precise altitude and the recent results
from BESS show only a weak dependence of the charge ratio on the altitude.
The BESS data indicate a 3\,\% difference between Tsukuba and Mt. 
Norikura which has an altitude of 2770\,m~\cite{bess_mount01}.

The disability of GHEISHA, the standard hadronic interaction model
in the detector simulation tool GEANT 3.21, in reproducing the data of
atmospheric muons surprises. But in fact serious deficits of GHEISHA
have already been proved in direct model tests. In 
Refs.~\cite{ferrari96,wentz99,wentz01} it has been reported that GHEISHA
violates the energy, 
momentum, charge and baryon number conservation in the single hadronic 
interaction.

At least the energy conservation is also violated on average as can
be shown by the simulation of extensive air showers with standard CORSIKA.
CORSIKA allows to summarize all the energy deposited in the atmosphere
during the shower development. Using GHEISHA as low energy hadronic
interaction model, an augmentation of energy of a complete shower is 
observed. This increase of energy is about 5\,\% at 10$^{15}$\,eV and
7\,\% at 10$^{14}$\,eV. Therefore, the GHEISHA version used in
GEANT3~\cite{geant} should not be 
used in any serious simulation of atmospheric neutrino fluxes. This holds 
especially for the neutrino flux calculations of Plyaskin,
which are based on GHEISHA, only.
After finishing the simulations, correction patches for GHEISHA became
available which improve essentially the energy 
conservation~\cite{gheisha_patch}.
 
\begin{figure}[!hb]
\includegraphics[width=8.cm]{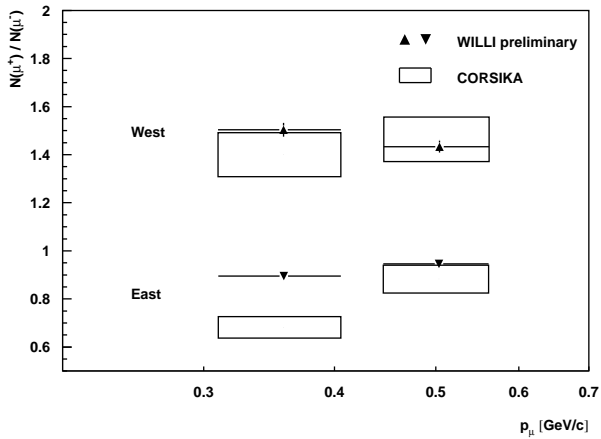}
\caption{\label{fig:ratio_ew} The East-West-effect of the muon charge
ratio as measured by the
WILLI detector in comparison with calculations of CORSIKA using DPMJET.
The detector acceptance of WILLI is taken into account by processing
the raw results of CORSIKA by the detector simulation program of WILLI.
East and WEST mean that the detector looks to the East and West, respectively.
}
\end{figure}

\subsection{The East-West-effect of the muon charge ratio}

Data for inclined muons allow a check of the calculations in the curved
geometry of the Earth. Using the so-called East-West-effect of the muon
charge ratio, caused by the influence of the geomagnetic field, the way
of handling the field in the calculation can be verified, too.

Fig.~\ref{fig:ratio_ew} shows preliminary results of the WILLI experiment
for muons observed in East and West direction having a mean zenith angle
of 35$^\circ$~\cite{willi_east_west} in comparison with CORSIKA simulations
on basis of DPMJET. The CORSIKA results are processed by a
full detector simulation of the experiment in order to account for the
complex acceptance of the instrument.

The agreement of the CORSIKA results with the strong East-West-effect observed
by the WILLI experiment, gives confidence that the corresponding effect
in the atmospheric neutrino flux is also handled well by CORSIKA.  

Muon data in various depths of the atmosphere would provide a further
possibility for the revision of calculations on atmospheric particle fluxes.
Unfortunately the rise and decline time of the actual balloon measurements
are such fast, that the corresponding muon data have large statistical
errors. Additionally the atmospheric pion flux causes systematic
errors in some instruments. While the pion flux on sea-level is only
0.5\,\% of the muon flux it reaches 50\,\% when approaching the top of
atmosphere. 

Nevertheless it has to be pointed out that the atmospheric muon flux,
in contrast to the neutrino flux where every produced neutrino reaches
ground level, is a highly differential quantity, because
most muons are already absorbed before reaching ground level. Therefore
possible differences, for example in the nuclear interaction models
are enhanced from one hadronic interaction to the next. Thus the calculation
of the ground level muon flux has higher theoretical uncertainties
than the calculation of atmospheric neutrino fluxes.

\section{Calculation of atmospheric neutrino fluxes}
\subsection{The vertical neutrino fluxes in Kamioka}

The calculation of atmospheric neutrino fluxes for Kamioka is split in
two separate calculations. The downward going neutrinos are simulated locally
for Kamioka, while the upward going neutrinos are calculated from primary
particles distributed over the entire Earth and only neutrinos passing
in a circle of 1000 km distance from Kamioka are used in the 
further analysis. 

\begin{figure}[!hb]
\includegraphics[width=8.5cm]{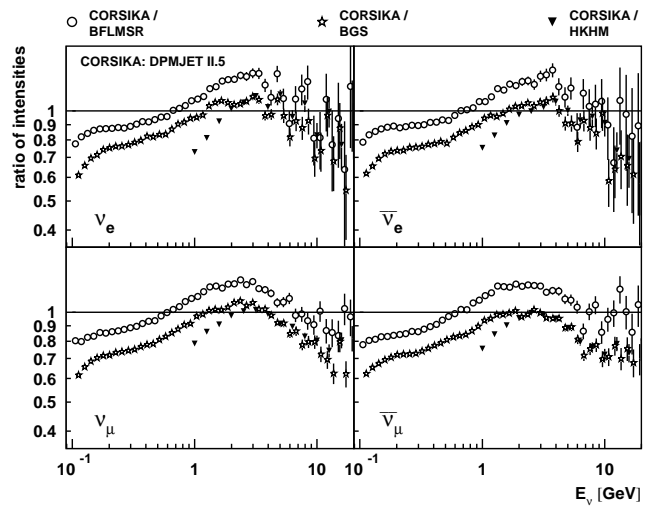}
\caption{\label{fig:diff_dpmjet} The vertical differential
intensities of the different neutrino flavors in Kamioka, displayed as the
ratio between the CORSIKA results using DPMJET as hadronic interaction
model and the calculations of BGS, HKHM and BFLMSR.}
\end{figure}

\begin{table*}
\caption{\label{tab:diff_fluxes}The vertical differential intensity $f(\nu)$
and the errors $\Delta(\nu)$ for downward going neutrinos in Kamioka
as calculated with CORSIKA using DPMJET and VENUS + UrQMD. The
fluxes and the corresponding statistical errors due to the limited number of
events in the Monte Carlo simulation are given in units of
(m$^2$\,s\,sr\,GeV)$^{-1}$.}
\begin{ruledtabular}
\begin{tabular}{c|cccccccc|cccccccc}
& \multicolumn{8}{c}{DPMJET} & \multicolumn{8}{c}{VENUS + UrQMD}\\
$ E_\nu$ $[$GeV$]$ & 
        $f(\nu_e)$ & $\Delta(\nu_e)$ & 
        $f(\overline{\nu_e})$ & $\Delta(\overline{\nu_e})$ &
	$f(\nu_\mu)$ & $\Delta(\nu_\mu)$ & 
	$f(\overline{\nu_\mu})$ & $\Delta(\overline{\nu_\mu})$ &
	$f(\nu_e)$ & $\Delta(\nu_e)$ &
	$f(\overline{\nu_e})$ & $\Delta(\overline{\nu_e})$ &
	$f(\nu_\mu)$ & $\Delta(\nu_\mu)$ &
	$f(\overline{\nu_\mu})$ & $\Delta(\overline{\nu_\mu})$\\
\hline
0.112&1303. & 6.1 &1251. & 5.9 &2708. & 8.8 & 2727.& 8.8  & 1341. & 6.4  & 1330.&6.4  &2838.&9.3  &2857.&9.4  \\
0.141&1142. & 5.1 &1100. & 5.0 &2336. & 7.2 & 2329.& 7.2  & 1154. & 5.3  & 1153.&5.3  &2430.&7.7  &2422.&7.7  \\
0.178&921.3 & 4.1 &875.5 & 4.0 &1894. & 5.8 & 1870.& 5.8  & 932.5 & 4.2  & 920.7&4.2  &1995.&6.2  &1985.&6.2  \\
0.224&702.1 & 3.2 &655.1 & 3.0 &1455. & 4.5 & 1432.& 4.5  & 703.0 & 3.3  & 678.3&3.2  &1526.&4.8  &1506.&4.8  \\
0.282&506.0 & 2.4 &473.3 & 2.3 &1075. & 3.5 & 1050.& 3.4  & 505.0 & 2.5  & 488.5&2.4  &1114.&3.7  &1094.&3.7  \\
0.355&361.6 & 1.8 &327.8 & 1.7 &775.8 & 2.6 & 755.8& 2.6  & 347.6 & 1.8  & 334.3&1.8  &776.9&2.7  &769.5&2.7  \\
0.447&247.8 & 1.3 &221.3 & 1.3 &542.1 & 2.0 & 526.3& 1.9  & 231.4 & 1.3  & 218.1&1.3  &528.1&2.0  &512.7&2.0  \\
0.562&164.1 & .96 &142.9 & .90 &371.7 & 1.4 & 358.8& 1.4  & 150.8 & .96  & 140.0&.93  &349.6&1.5  &340.3&1.4  \\
0.708&106.8 & .69 &92.10 & .64 &246.3 & 1.1 & 234.3& 1.0  & 94.37 & .68  & 85.94&.65  &224.2&1.0  &212.4&1.0  \\
0.891&66.74 & .49 &56.19 & .45 &160.2 & .76 & 149.0& .73  & 57.09 & .47  & 50.84&.44  &140.4&.74  &134.1&.72  \\
1.122&39.37 & .33 &33.05 & .31 &99.78 & .53 & 92.37& .51  & 32.49 & .32  & 29.51&.30  &84.70&.51  &80.70&.50  \\
1.413&23.33 & .23 &19.20 & .21 &59.89 & .37 & 54.97& .35  & 18.74 & .21  & 16.47&.20  &50.29&.35  &46.93&.34  \\
1.778&12.89 & .15 &10.22 & .14 &33.97 & .25 & 30.76& .23  & 10.21 & .14  & 8.636&.13  &28.59&.24  &26.40&.23  \\
2.239&6.746 & .098&5.366 & .087&19.23 & .17 & 16.50& .15  & 5.345 & .091 & 4.579&.084 &15.73&.16  &14.30&.15  \\
2.818&3.413 & .062&2.632 & .054&10.24 & .11 & 8.821& .10  & 2.609 & .056 & 2.270&.053 &8.783&.10  &7.615&.096 \\
3.548&1.611 & .038&1.347 & .035&5.236 & .068& 4.432& .063 & 1.258 & .035 & 1.115&.033 &4.645&.067 &3.983&.062 \\
4.467&.741 & .023 &.583 & .020 &2.566 & .043& 2.168& .039 & .6162 & .022 & .5615&.021 &2.490&.044 &2.106&.040 \\
5.623&.299 & .013 &.241 & .012 &1.266 & .027& 1.044& .024 & .3047 & .014 & .2447&.012 &1.269&.028 &1.065&.026 \\
7.079&.133 & .0077&.117 & .0073&.6337 & .017& .4648& .014 & .1278 & .0079& .0991&.0069&.5997&.017 &.5384&.016 \\
8.913&.060 & .0046&.049 & .0042&.2966 & .010& .2328& .0091& .0749 & .0054& .0452&.0042&.3104&.011 &.2320&.0095\\
11.22&.023 & .0026&.016 & .0022&.1504 & .0065&.1210& .0059& .02913& .0030& .0187&.0024&.1674&.0072&.1168&.0060\\

\end{tabular}
\end{ruledtabular}
\end{table*}

\begin{figure}[!hb]
\includegraphics[width=8.5cm]{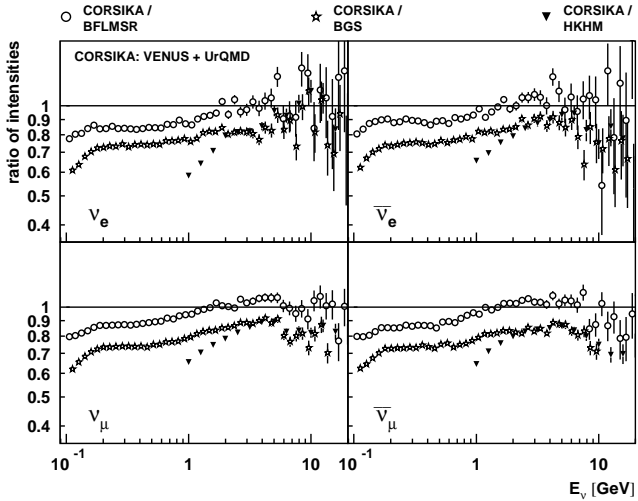}
\caption{\label{fig:diff_urqmd} The vertical differential
intensities of the different neutrino flavors in Kamioka. Shown is the
ratio between the CORSIKA results 
using VENUS + UrQMD as hadronic interaction model
and the calculations of BGS, HKHM and BFLMSR.}
\end{figure}

This procedure causes a large difference in the number of primary
particles needed in the simulation for obtaining the same statistical
accuracy for the up- and downward going fluxes. In the present
simulation the number of upward going neutrinos is still a factor of 8
smaller.

Tab.~\ref{tab:diff_fluxes} gives the differential intensities for vertical
neutrinos obtained with CORSIKA, using DPMJET and VENUS + UrQMD, respectively.
In Figs.~\ref{fig:diff_dpmjet} and \ref{fig:diff_urqmd} the results are
compared directly with the calculations of BGS, HKHM and BFLMSR.

The inclusive neutrino fluxes obtained with CORSIKA are evidently lower
than the fluxes given by BGS and HKHM. The differential fluxes at 0.1\,GeV are
about 40\,\% smaller than the BGS fluxes and become comparable at energies 
in the GeV range. 
The agreement of the CORSIKA results using DPMJET and using VENUS + UrQMD with 
the BFLMSR calculation is better. The deviation of these absolute flux 
calculations over practically the whole energy range remains less than
20\,\%. The energy dependence of the neutrino fluxes between BFLMSR and
VENUS + UrQMD is quite similar while DPMJET shows a systematic difference
to BFLMSR.

\begin{figure}[!ht]
\includegraphics[width=8.5cm]{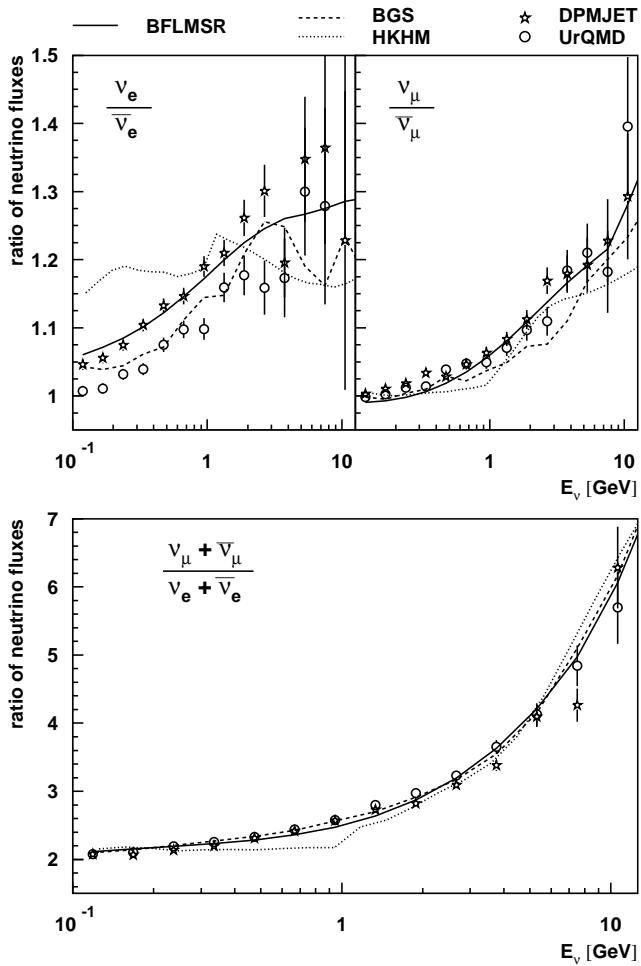}
\caption{\label{fig:vert_ratio} The ratio between different neutrino
flavors in the vertical flux in Kamioka as calculated by CORSIKA with
DPMJET and with VENUS + UrQMD.}
\end{figure}

Fig.~\ref{fig:vert_ratio} displays the ratio between the different neutrino
flavors in the vertical downward going flux. The agreement of all calculations
for the ratio of muon neutrinos to electron neutrinos is very good. The deviation
of the HKHM results and the discontinuity at $E_\nu\,=\,1$\,GeV are caused
by different approaches in the model. Below 1\,GeV the values of HKHM are
averaged over the zenith angle, only above 1\,GeV they stand for
vertical, downward going neutrinos.
For energies below 3\,GeV the differences between the other models
are on the level of 2\,\% or better.

Some differences between the calculations are observed in the 
ratio of muon neutrinos to muon antineutrinos. The 
CORSIKA calculations with DPMJET and VENUS + UrQMD,
and the BFLMSR calculations agree perfectly. The calculations
of BGS predict a lower ratio at 3\,GeV while the calculations of
HKHM are different around 1\,GeV and show a smaller rise of the ratio 
at high energies. 

The ratio of electron
neutrinos to electron antineutrinos reveals larger differences. The results
of HKHM behave quite different from the results of all other models.
Interestingly, DPMJET results agree with BFLMSR results, while 
VENUS + UrQMD results agree with BGS results. Due to the close correlation
between the ratio of electron neutrinos to electron antineutrinos and the 
charge ratio of muons, these findings allow to rule out the results of
VENUS + UrQMD in this particular quantity, meaning that the results of BGS
are suspicious in this aspect, too.

An interesting quest for the CORSIKA calculations with their inclusion 
of the precise geometry of the Earth, are natural differences 
between the up- and downward going neutrino fluxes in Kamioka.
Such differences could contribute to the measured asymmetry, which
is commonly attributed to the oscillation of neutrinos. Any natural
difference based on the geographical environment has a direct impact on
the analysis of the neutrino oscillations and changes finally the obtained
oscillation parameters. 

\begin{figure}[!ht]
\includegraphics[width=8.cm]{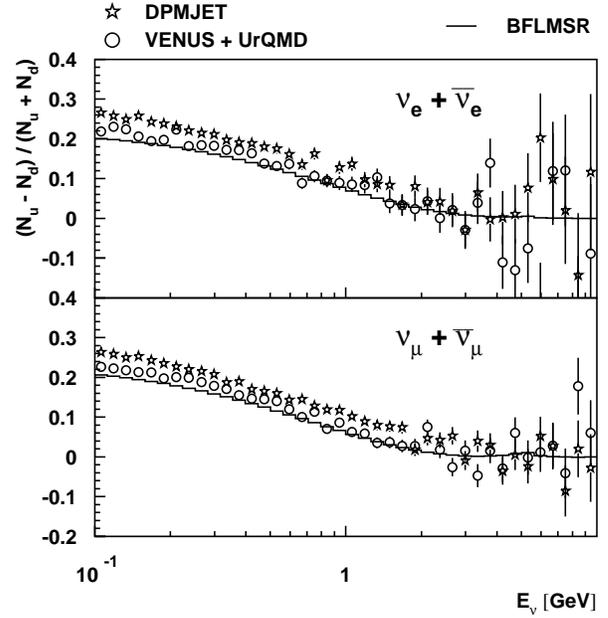}
\caption{\label{fig:up_down} The asymmetry of the up- ($N_u$) and
downward ($N_d$) going neutrino fluxes for Kamioka as calculated
by CORSIKA using DPMJET and VENUS + UrQMD in comparison with the
calculation of BFLMSR. The asymmetry is expressed by the ratio
between the difference and the sum of the up- and downward going 
neutrino fluxes.
}
\end{figure}

A major difference between Kamioka and its antipode in the 
South Atlantic comes from the geomagnetic cut-off. While the vertical
cut-off in Kamioka is 12.3\,GV, the South Atlantic region is influenced
by the so-called South Atlantic magnetic field anomaly leading to a
vertical cut-off at the antipode of only 8.6\,GV. This causes an
asymmetry between the intensities of up- and downward going neutrinos
for Kamioka, as can be seen in Fig.~\ref{fig:up_down}.

The asymmetry of 20\,\% having been observed in the calculations 
of BFLMSR represents the raw effect based on the differences in the
geomagnetic cut-off, because the calculation does not include any local
magnetic field. In the CORSIKA simulations the local field and an additional 
contribution to the up-down asymmetry, caused by the different
elevation of the surface over sea-level in Kamioka and in the
antipode region in the South Atlantic, are taken into account. 

\begin{figure*}[!ht]
\vspace{0.3cm}
\includegraphics[width=17.9cm]{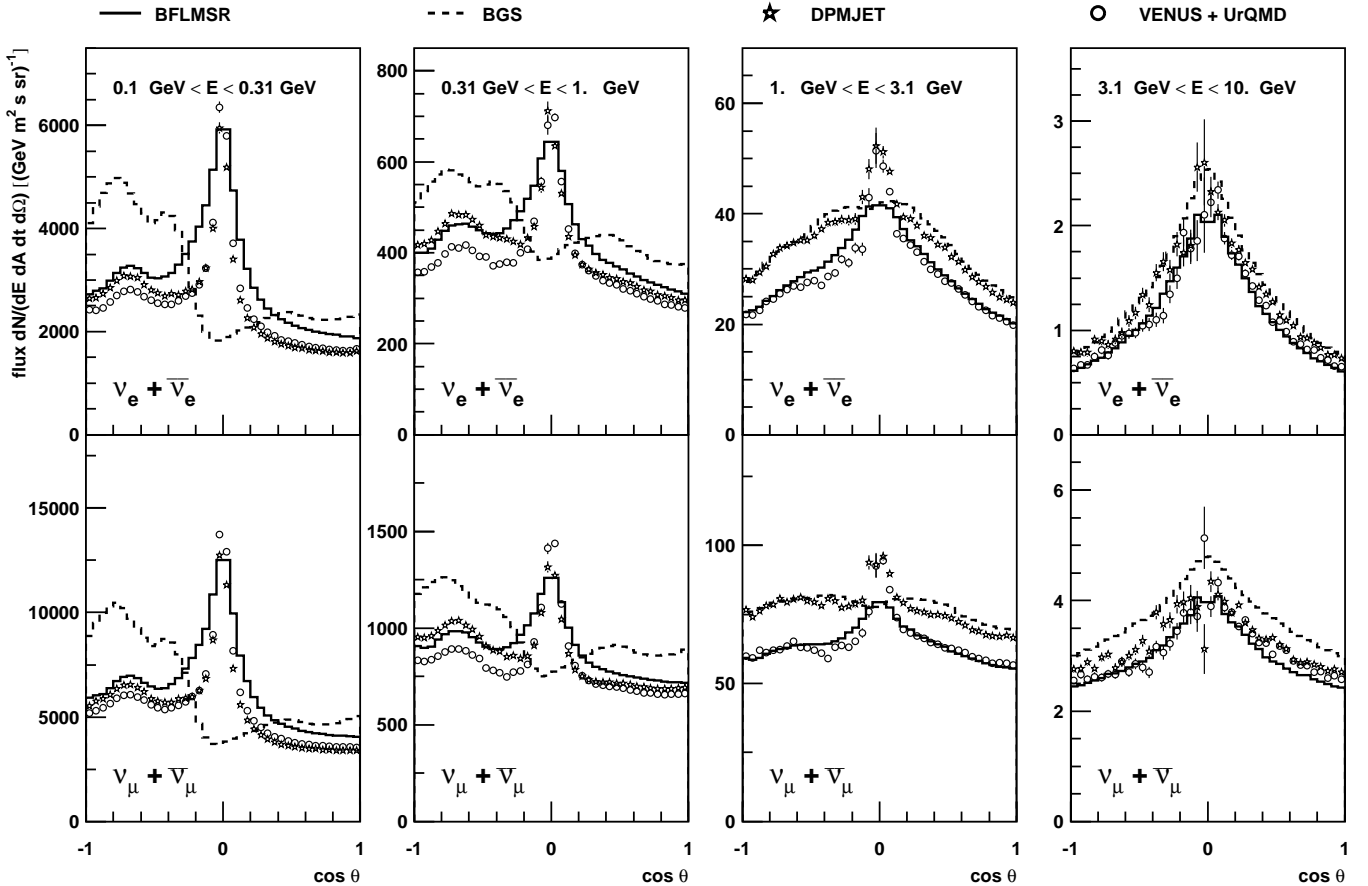}
\caption{\label{fig:neut_theta} The zenith angle dependence of
the neutrino intensities in Kamioka as calculated by CORSIKA with DPMJET
and with VENUS + UrQMD in comparison with the calculations of
BFLMSR and BGS. The first column shows the plots for the energy
interval 0.1 to 0.31\,GeV, the second for 0.31 to 1\,GeV, the
third for 1 to 3.1\,GeV, and the last for 3.1 to 10\,GeV.}
\end{figure*}

The location of the Super-Kamiokande detector in the mountains causes
an altitude difference of several hundred meters compared to the
average altitude of the antipode region. Thus
in the South Atlantic the shower development is longer and more 
neutrinos are produced in the shower. 
Further details on the influence of the local magnetic field
and the geomagnetic cut-off are investigated in Sec.~\ref{sec:cutandfield}.
The effect of the contrary seasons in Japan and the South Atlantic,
which is taken into account by using the appropriate atmospheric models
does not lead to any observable effect, the effect is smaller than the
actual statistical errors.

\begin{figure}[!ht]
\includegraphics[width=8.5cm]{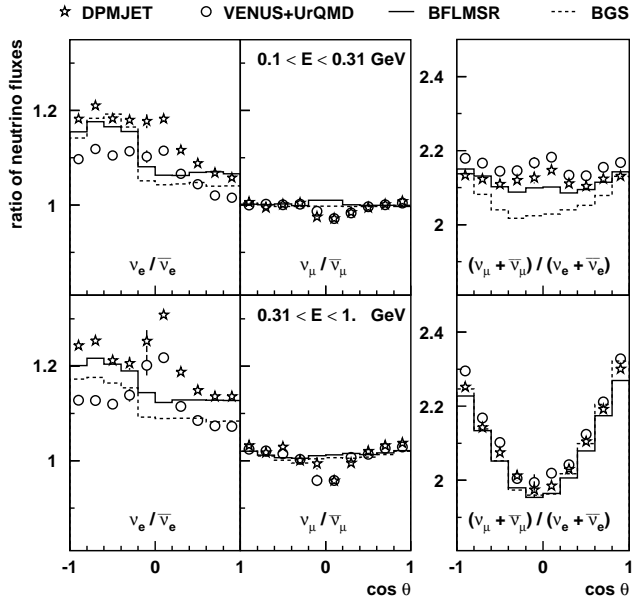}
\caption{\label{fig:rat_theta} The results of CORSIKA
using DPMJET and VENUS + UrQMD for the zenith angle dependence of
the different ratios between the neutrino flavors.
The values are compared with the calculations of
BFLMSR and BGS. The plots in the first row are for the energy interval
0.1 to 0.31\,GeV and in the second row for 0.31 to 1\,GeV. 
If the error bar is not drawn, the error is smaller than the symbol size.
}
\end{figure}

\begin{figure*}[!ht]
\vspace{0.5cm}
\includegraphics[width=17.9cm]{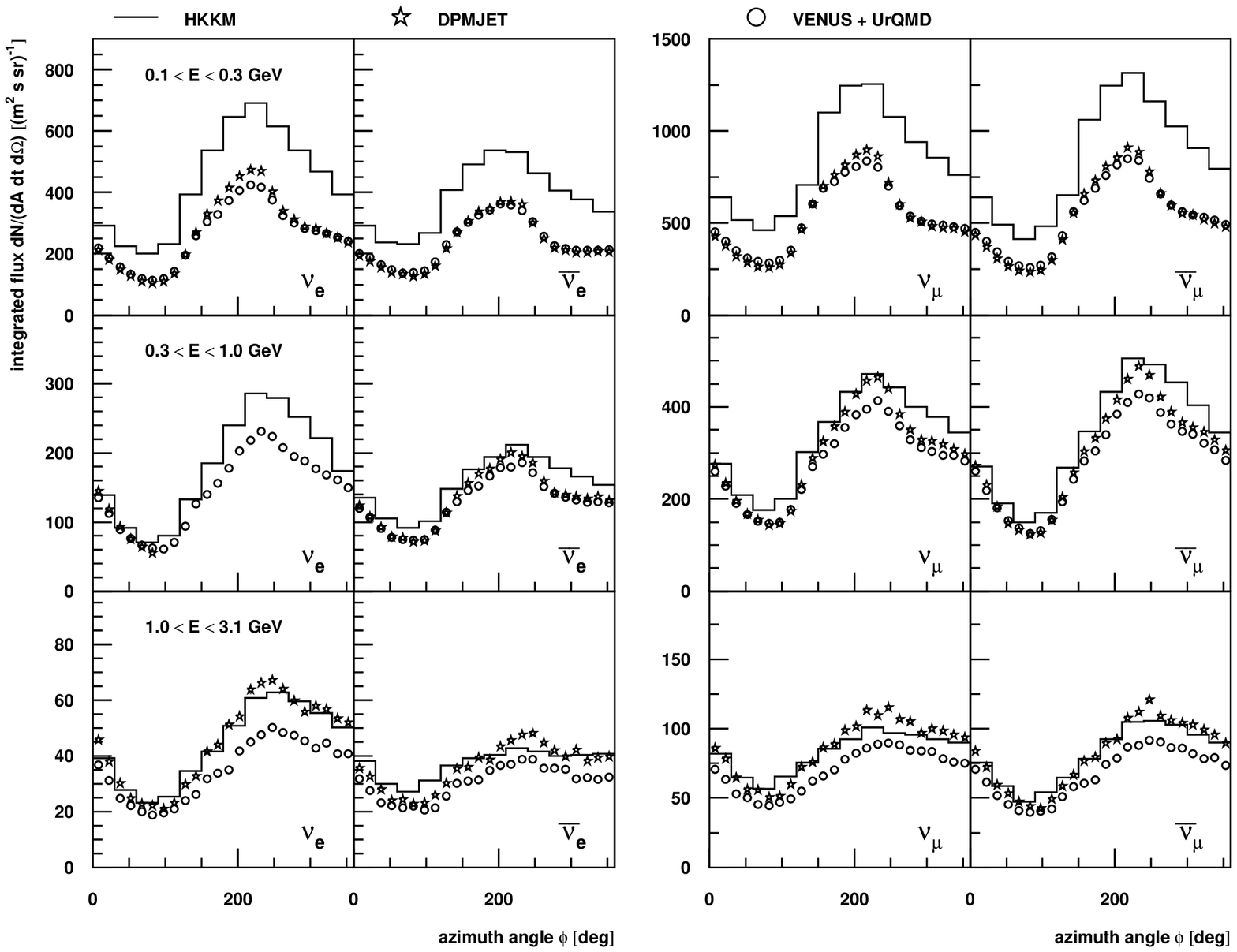}
\caption{\label{fig:flux_phi_honda} The azimuth angle dependence of
the neutrino fluxes in Kamioka as calculated with CORSIKA for different energies
and neutrino flavors compared with the calculations of HKKM. The data
in the first row are integrated in an energy interval from 0.1 to 0.3\,GeV,
in the second row from 0.3 to 1\,GeV, and in the third row 
from 1 to 3.1\,GeV.
The neutrinos selected are from
both hemispheres and have a $|\cos \theta| > 0.5$.
In this diagram $\phi = 0$ indicates a particle going in 
the magnetic North direction. The errors are smaller than the symbol
sizes.
} 
\end{figure*}

\subsection{The directional dependence of the neutrino fluxes in Kamioka}

The dependence of the neutrino fluxes on the zenith angle is shown
in Fig.~\ref{fig:neut_theta}. The three
dimensional calculations of BFLMSR and CORSIKA show an
enhancement of the neutrino fluxes near the horizon. This enhancement
is based on a geometrical effect, i.e. the spherical shell 
geometry of the neutrino production volume~\cite{lipari_3d00}.   
This effect has been neglected in all one dimensional simulations like
HKHM and BGS. The strength of the effect shows clearly the necessity
of the time consuming three dimensional simulations in a spherical geometry.
The agreement of the calculation with VENUS + UrQMD and with BFLMSR is again
better, while the DPMJET results show systematically higher fluxes for
energies between 1 and 3\,GeV. 

The dependence of the resulting ratio between muon neutrinos and electron 
neutrinos on the zenith angle is shown in Fig.~\ref{fig:rat_theta}.
Only the results for energies below 1\,GeV are plotted, for higher
energies no difference between all the four calculations is observed.
As in the case of the ratios between vertical neutrino fluxes the largest
differences are observed in the ratio of electron neutrinos to electron
antineutrinos. The CORSIKA results show a strong increase of the ratio
near the horizon. The origin of this effect will be investigated in 
Sec.~\ref{sec:cutandfield}.

Also a 8\,\% difference of the ratio of muon neutrinos to electron
neutrinos at low energies can be observed near the horizon. The results of
the BGS calculation lead to very low values for this quantity, 
and may be an artifact of the calculation in a one dimensional geometry.
  
The dependence of the neutrino fluxes on the azimuth angle is shown
in Fig.~\ref{fig:flux_phi_honda}. The agreement between the calculations
with DPMJET and with VENUS + UrQMD for westward going neutrinos is very
good, but for eastward going neutrinos some noticeable differences 
are observed at higher energies. This is a secondary effect of the difference
in the momentum spectra of the reaction products between both models,
but it displays also an instructive example how the interaction model
influences results which are commonly assumed to have a geometrical
nature. 

The detailed comparison with the results of the HKKM calculation
shows a very good agreement in the shape of the azimuthal distribution.
At lowest energies the HKKM calculation leads to much higher fluxes.
The authors state this overestimation to be caused by the use of the
old COSMOS interaction models. A new calculation using DPMJET as hadronic
interaction model will overcome this problem. 

\begin{figure}[!ht]
\vspace{0.5cm}
\includegraphics[width=8.5cm]{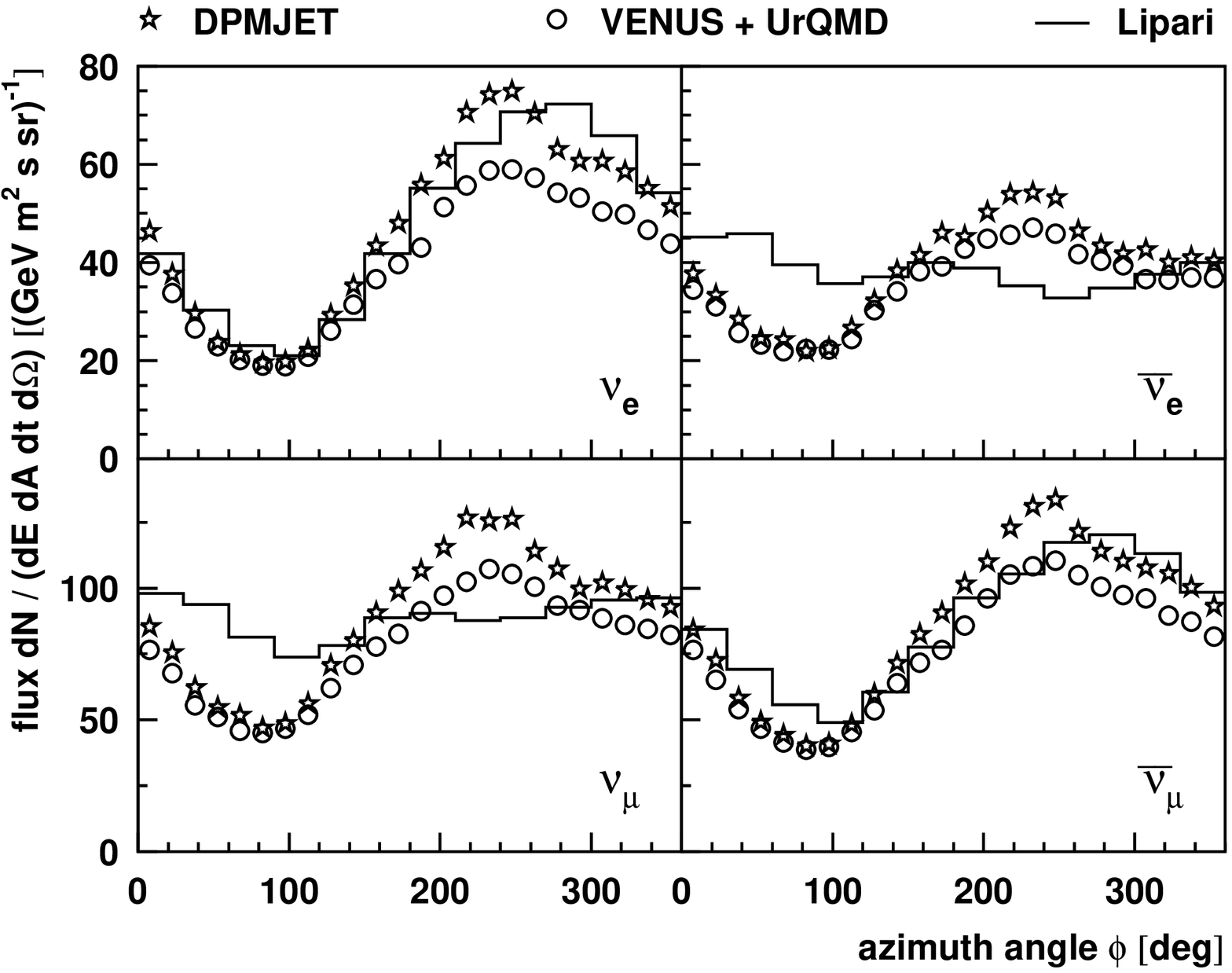}
\caption{\label{fig:flux_phi} The azimuth angle dependence of
the neutrino fluxes as calculated by CORSIKA with DPMJET
and with VENUS + UrQMD, compared with a calculation of
Lipari et al.~\cite{lipari_geo98,lipari_3d00}. The neutrinos used
in the analysis result from
both hemispheres, requesting $|\cos \theta| > 0.5$ and an energy between
0.5 and 3\,GeV. The calculations of Lipari have been normalized to
the fluxes obtained with DPMJET. In this
diagram $\phi = 0$ indicates a particle going towards
the magnetic North direction. The errors are smaller than the symbol
sizes.} 
\end{figure}

The good agreement between CORSIKA results and the calculation of HKKM
in the azimuthal distribution is by far not trivial, as shows the comparison
of the CORSIKA results with calculations of 
Lipari et al.~\cite{lipari_geo98,lipari_3d00} in
Fig.~\ref{fig:flux_phi}. 
Here the shapes of the distributions for electron neutrinos and 
muon antineutrinos are compatible, but strong disagreement exists
for electron antineutrinos and muon neutrinos.

\begin{figure}[!ht]
\vspace{0.5cm}
\includegraphics[width=8.5cm]{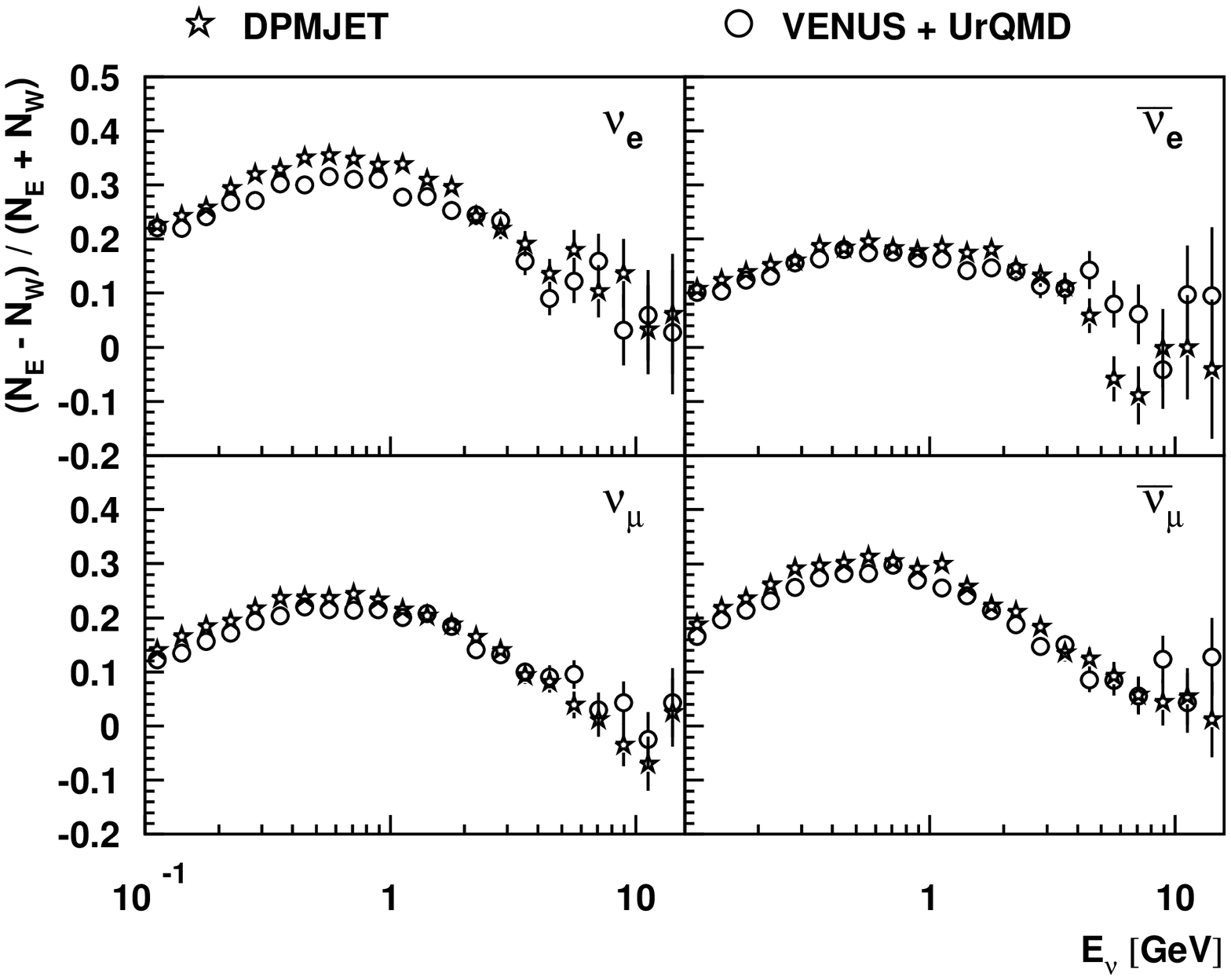}
\caption{\label{fig:flux_phi_en} The energy dependence of the
East-West-asymmetry in the atmospheric neutrino flux as calculated by
CORSIKA with DPMJET and with VENUS + UrQMD.} 
\end{figure}

\begin{figure}[!ht]
\vspace{0.5cm}
\includegraphics[width=8.5cm]{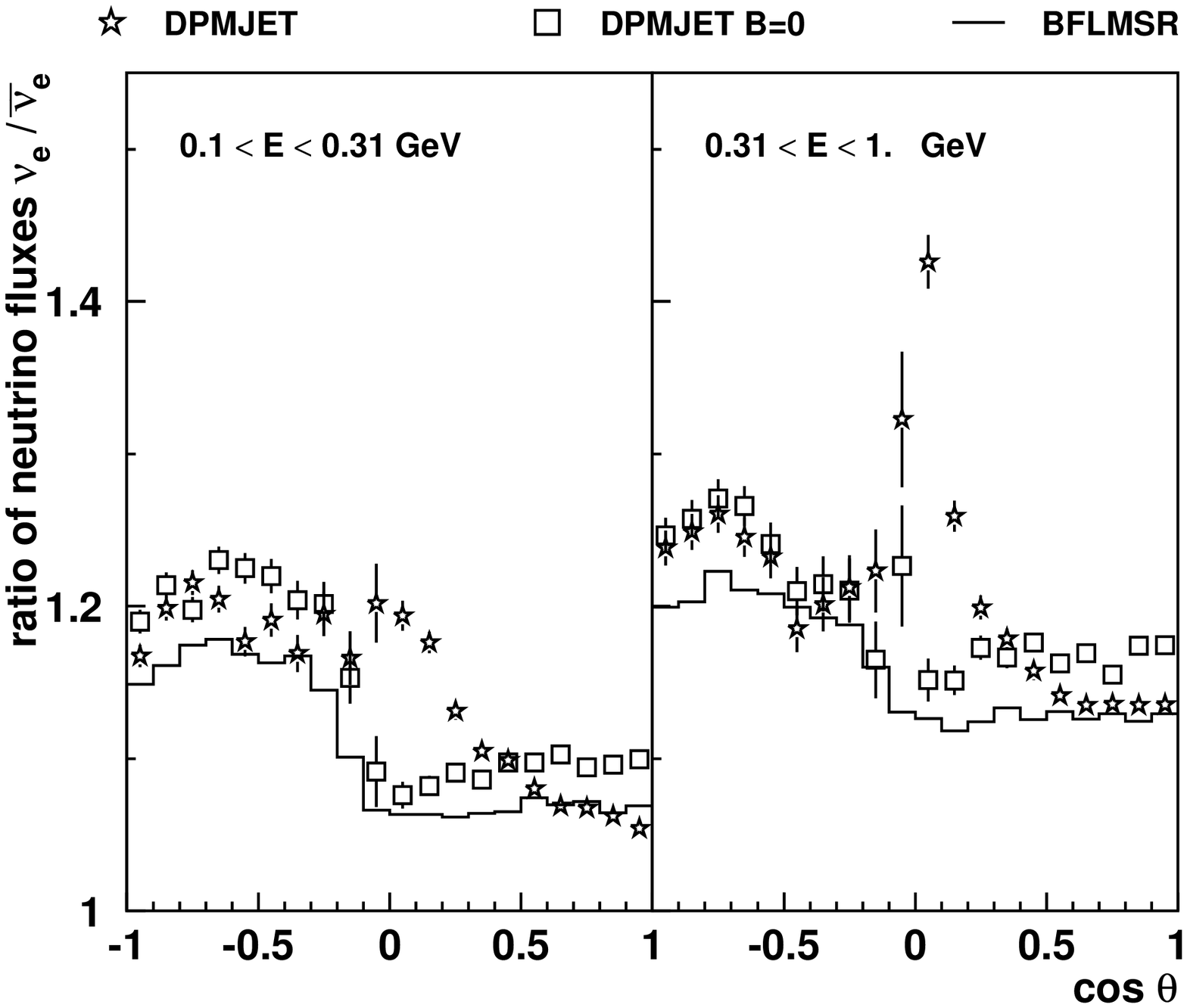}
\caption{\label{fig:rat_theta_0b} Zenith angle dependence of the 
ratio between electron neutrino and electron antineutrino fluxes 
in Kamioka. The results of CORSIKA with DPMJET considering the 
local magnetic field are compared with results of CORSIKA and  
BFLMSR neglecting the local magnetic field.} 
\end{figure}

The results can be expressed by the East-West-asymmetry
$A_{EW} = (N_E - N_W) / (N_E + N_W)$, where $N_E$ and $N_W$ stand for the
particle fluxes of neutrinos going to the East and West, respectively.
Fig.~\ref{fig:flux_phi_en} shows the energy dependence of the
East-West-asymmetry. Again the CORSIKA results with DPMJET
have a slightly higher asymmetry than the calculations with VENUS + UrQMD. 
The distributions of all neutrino flavors show similar shapes. 
The strongest asymmetry is observed for electron neutrinos and the weakest 
for electron antineutrinos. All neutrino flavors exhibit a maximal asymmetry
for an energy around 800\,MeV.

\begin{figure*}[!ht]
\vspace{0.2cm}
\includegraphics[width=17.9cm]{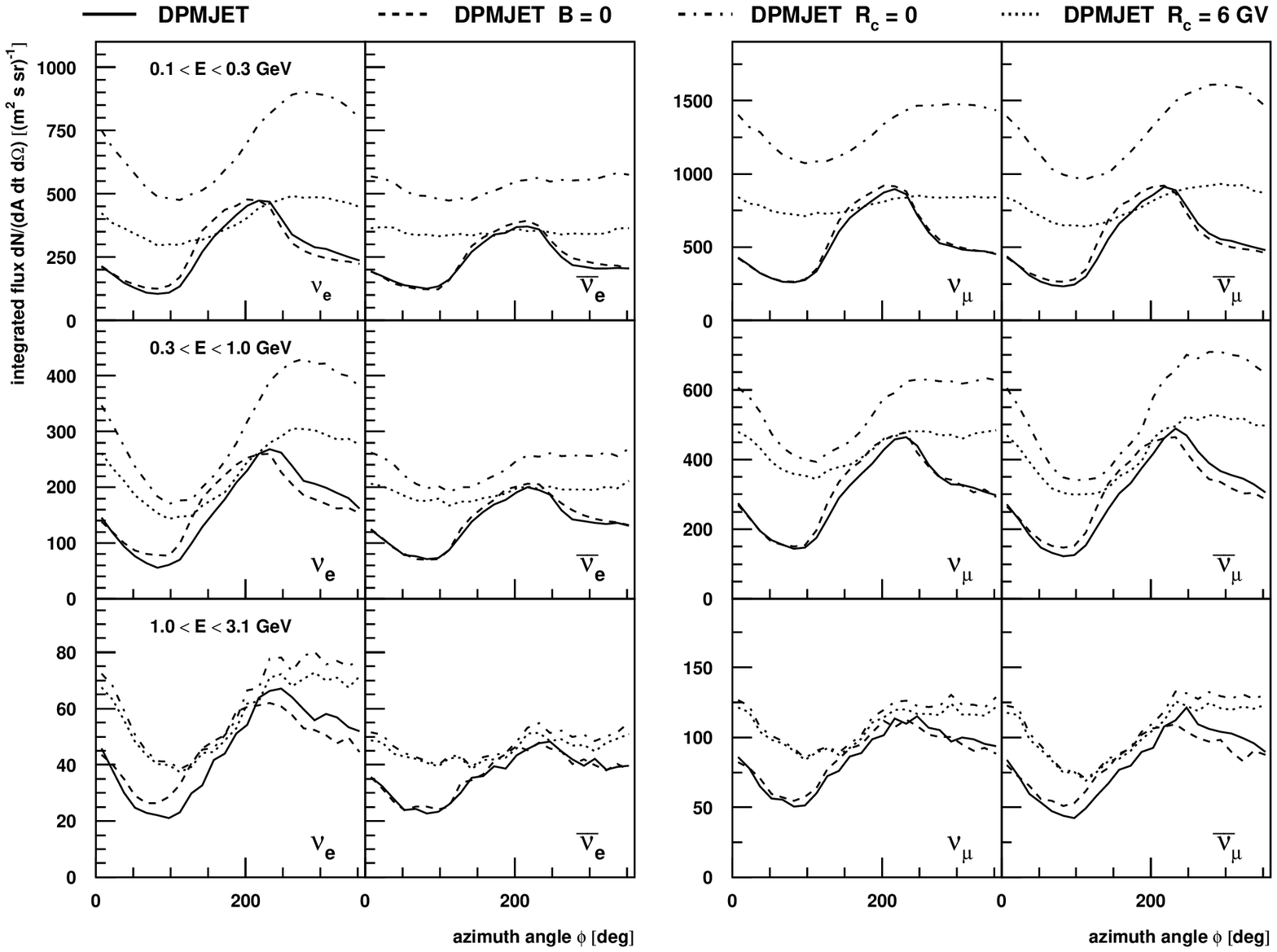}
\caption{\label{fig:flux_phi_0boc} The azimuth angle dependence of
the atmospheric neutrino fluxes in Kamioka as calculated by CORSIKA under
different conditions. The neutrinos used in the analysis result from
both hemispheres and have a $|\cos \theta| > 0.5$.
Beside the results described already above,
the azimuthal distributions are shown for a zero local magnetic field, a zero
geomagnetic cut-off and an isotropic cut-off of 6\,GV. In this
diagram $\phi = 0$ indicates a particle going towards
the magnetic North direction. The distributions in the first row are
integrated over an energy interval between 0.1 and 0.3\,GeV, in the second 
between 0.3 and 1\,GeV, and in the third between 1 and 3.1\,GeV.} 
\end{figure*}

\subsection{\label{sec:cutandfield}The influences of the geomagnetic cut-off
and the local magnetic field}

In order to investigate the individual influences of the geomagnetic cut-off
and of the local magnetic field, the calculations of the atmospheric 
neutrino fluxes for Kamioka with DPMJET have been repeated twice under
the same conditions, except setting once the local magnetic field and 
once the geomagnetic cut-off to zero. This procedure allows to
disentangle the individual influences of the two effects.

Due to the fact that charged particles do not win or loose energy in a 
magnetic field, the influence of the local magnetic field
on the total neutrino fluxes is negligible. The main effects are expected
in the ratios of neutrinos and in the azimuthal distribution of the fluxes.
Especially the ratio of electron neutrinos to electron antineutrinos shows
a strong effect because the electron neutrinos are 
predominantly produced by positive muons and the electron antineutrinos 
by negative muons. 

Muon neutrinos and muon antineutrinos are produced also in the decay of
charged pions. In contrast to the muon decay, muon neutrinos result here
from the decay of positive and muon antineutrinos from the decay of negative 
particles. 
Due to the shorter life time and the higher momentum the total
bending of pions is less and the bending of the muons is preponderating, 
but the total effect of the local magnetic field on the
muon neutrinos remains weaker.

The effect of the inclusion of the local magnetic field in the calculation
is shown in Fig.~\ref{fig:rat_theta_0b}. The increase of the ratio between
electron neutrinos and electron antineutrinos near the horizon as observed
in Fig.~\ref{fig:rat_theta} has to be attributed completely to the bending
of the charged shower particles in the atmosphere.

The CORSIKA results for the azimuthal dependence of the atmospheric neutrino 
fluxes under the different conditions are displayed 
in  Fig.~\ref{fig:flux_phi_0boc}. The differences are pronounced
for smaller energies. At higher energies all the different conditions
lead to identical fluxes. The influence on the shape of the azimuthal
distribution is weak, but only for detector sites
with a high geomagnetic cut-off.

Without consideration of the geomagnetic cut-off, much higher neutrino
fluxes are obtained due to the higher fluxes of primary particles. The
asymmetry in the azimuthal distribution results here only from the deflection
of charged shower particles in the local magnetic field. The characteristics
of this asymmetry is very similar to the East-West-effect caused by the
geomagnetic cut-off, a consequence of the excess of positive particles
in the atmosphere, on which the magnetic field acts in a similar way as
on the primary proton flux. This argument is supported by the
different behavior of electron antineutrinos, which are produced only in
the decay of negative muons.

In order to illustrate the transition between a zero and a
high geomagnetic cut-off, the results of a calculation assuming an isotropic
cut-off of 6\,GV have been added also in Fig.~\ref{fig:flux_phi_0boc}. 
These results show that a neglect of the local magnetic field, as it is done
in many calculations of atmospheric neutrino fluxes, may lead to wrong
azimuthal distributions at least for detector sites with a comparable
low geomagnetic cut-off.  

\section{Conclusion}

This work aims at a new procedure for the calculation of atmospheric
neutrino fluxes with considering various influences which have not been taken
into account so far or, if ever, only in a less rigorous way. The capabilities of
the procedure are demonstrated by a particular calculation of the detailed neutrino
fluxes in Kamioka. The detailed procedure applies the air shower simulation code
CORSIKA in the version 6.000, which has been extended and modified for a reliable
simulation of the cascading interactions induced in the atmosphere by
primary particles of the low energy part of the cosmic ray spectrum.

A description of the solar modulation and tables for the geomagnetic 
cut-off, calculated in a detailed Monte Carlo simulation, have been introduced. 
In addition, for the first time for atmospheric neutrino flux calculations, 
the geography of the Earth is taken into account by a digital elevation model,
tables for the local magnetic field in the atmosphere, and
various atmospheric models for different climatic zones and seasons.
These extensions are not yet part of the standard CORSIKA
package.

CORSIKA features a precise particle tracking, including
the deflection of the charged shower particles in the local magnetic
field, the energy loss by ionization and multiple scattering. The
used primary flux is based on the recent measurements of the prototype
of the AMS-experiment. These data allow also a test of the calculations
for the geomagnetic cut-off.

An important aspect of the calculations is the question of the adequate
hadronic interaction model used as generator of the flux calculations. 
This question is approached by using the possibilities of the CORSIKA
code to operate optionally with various different hadronic interaction models.
The models are scrutinized by an extensive comparison with measured fluxes
and charge ratios of atmospheric muons in different locations.

It turns out that the GHEISHA model leads to significant discrepancies 
with data from various experiments and predictions based on GHEISHA have
to be considered as highly doubtful. The use of DPMJET II.5 as well as of the 
combination VENUS + UrQMD results in differential muon fluxes which are 
in good agreement with the measurements. The DPMJET model reproduces  
the charge ratio of muons of vertical incidence, while the values obtained 
with VENUS + UrQMD appear systematically too small. The calculations with 
DPMJET agree also well with the preliminary results of the WILLI 
experiment for the East-West-effect of the charge ratio of muons with 
inclined incidence.

Subsequently CORSIKA is used with the described refinements to calculate 
the fluxes of atmospheric neutrinos for Kamioka. The resulting absolute neutrino
intensities are lower than those found in the classical calculations of BGS and HKHM,
but they are in good agreement with the recent three dimensional calculations
of BFLMSR. Using VENUS + UrQMD the deviations from BFLMSR predictions are
smaller than 20\,\% over the whole energy range and the overall energy
dependence is very similar. 

DPMJET leads to absolute fluxes, being also very similar to the simulations
of BFLMSR, but the energy dependence turns out to be slightly different. 
Nevertheless the better agreement of the DPMJET predictions with the measured
fluxes and charge ratios of atmospheric muons provides stringent
arguments in favor of this particular model.

The ratio of muon neutrinos to electron neutrinos and the ratio of muon neutrinos
to muon antineutrinos in the vertical downward flux are identical within
the statistical uncertainties for the CORSIKA calculations invoking DPMJET, 
VENUS + UrQMD, and the calculations of BFLMSR. But for lowest energy neutrinos
with horizontal incidence, the ratios between muon neutrinos and electron
neutrinos obtained with DPMJET and with VENUS + UrQMD are higher.

Significant differences are observed for the ratio of electron neutrinos to
electron antineutrinos. The DPMJET results for vertical neutrinos for 
this quantity agree with the results of BFLMSR, and the results of VENUS + UrQMD
agree with the results of BGS. Again the very good agreement in the correlated
quantity of the muon charge ratio gives a strong argument for DPMJET.
For horizontal neutrinos the CORSIKA results predict a strong increase
of the ratio at low energies.

The actual results have relevance for the analysis of the atmospheric
neutrino anomaly. Any change in the ratio of muon neutrinos to electron
neutrinos leads directly to a change of the oscillation parameters.   
Also the discrepancies found in the ratio of electron neutrinos to electron
antineutrinos are of particular interest for Super-Kamiokande, because    
the detection cross-sections for neutrinos are about three times larger than
for antineutrinos and it is not possible to distinguish between them in the 
experiment.

To quantify the influence of these effects on the neutrino oscillation
parameters would request a full detector simulation of the Super-Kamiokande
experiment based on the presented fluxes, a task which is beyond the scope
of this communication. It can be stated that the difference of
the neutrino fluxes presented here to those used in the oscillation 
analysis is not large enough to affect the claim of existence of neutrino 
oscillations from atmospheric neutrinos.

The use of two different hadronic interaction models, both of good repute
in interpretation of accelerator experiments, shows clearly the potential
influence of the hadronic interaction model
on the interpretation of the atmospheric neutrino anomaly.  
Due to the high quality of the recent measurements of the primary particle
fluxes, the main source of remaining uncertainties in the atmospheric flux
calculations has to be attributed now to the actual uncertainties in the
hadronic interaction models.

For studying the influence of the geomagnetic field and the origin of the
East-West-effect in the atmospheric neutrino flux, CORSIKA calculations with 
DPMJET, setting the local magnetic field to zero or skipping the geomagnetic
cut-off have been performed. The main influence of the local magnetic
field is found for the ratio of electron neutrinos to electron antineutrinos.
CORSIKA predicts for the first time a strong increase of the ratio
near the horizon.

The local magnetic field proves to be of minor
influence on the azimuthal distribution of neutrinos in Kamioka, and the
East-West-effect arises mainly from the azimuthal dependence of 
the primary particle flux caused by the geomagnetic cut-off rigidity.
The simulations without a geomagnetic cut-off show that this observation
is valid only for Kamioka with its relative high geomagnetic cut-off value.
For a neutrino detector site like Sudbury in Canada, where the vertical
geomagnetic cut-off is only 1.1\,GV, a measurable East-West-effect would 
originate exclusively from the bending of the charged shower particles in the local
magnetic field.

To which extent the Earth's geography significantly affects the
results of the calculations has not been investigated in detail by separate
calculations. The higher asymmetry of the up- and downward going particle 
fluxes, found in the actual calculations in comparison to results of BFLMSR, 
indicates an influence of the digital elevation model in the order of a few
percent. Compared to the
changes of the atmospheric depth by the different altitudes, the variation
induced by the different atmospheric models is small. The influence on the 
particle fluxes in Kamioka should be negligible.
Nevertheless for detector sites with extreme atmospheric conditions, like
the South Pole, the profile of the atmosphere may lead to noticeable
seasonal effects. 

\begin{acknowledgments}

This work has been supported by the Deutsche Forschungsgemeinschaft 
with the grants WE 2426/1-1 and WE 2426/1-2. The help of 
the International Bureau Bonn supporting the personal exchange
and of the Volkswagen-Stiftung for sponsoring valuable devices is
gratefully acknowledged. The authors are deeply indebted to
G. Schatz and H. Bl\"{u}mer which enabled and supported the
major part of this study in the Forschungszentrum Karlsruhe. 
The suggestions and valuable advice of H. St\"{o}cker and M. Bleicher
when incorporating the UrQMD model in the CORSIKA code and of
J. Ranft when applying DPMJET at low energies, are
appreciated. We thank R. Engel for carefully reading the manuscript
and T.K. Gaisser for providing us with tables of the BGS results.
One of the authors (J.W.) is grateful to the European Commission
Centre of Excellence (IDRANAP) in Bucharest for the grant which 
allowed the finalizing of this paper. 

\end{acknowledgments}

\bibliography{paper}

\end{document}